\documentclass{elsart}

\usepackage{natbib,amsmath,amsbsy,amsfonts,amscd}
\usepackage{vmargin}
\usepackage{latexsym}
\usepackage{graphicx}
\usepackage{amssymb}

\newtheorem{lemme}{Lemma}[section]

\newtheorem{Theorem}{Theorem}[section]

\newtheorem{Proposition}{Proposition}[section] 
\newtheorem{Remark}{Remark}[section]

\newcommand{\Ind}{\ifmmode{{\rm 1} \hskip -3pt {\rm I}}
    \else{\hbox{$1\hskip -3pt {\rm I}$}}\fi}
\newcommand{\g}{\nabla}
\newcommand{\Om}{\Omega}
\newcommand{\G}{\Gamma}
\newcommand{ \vit}{\hbox{\bf u}}

\newcommand{ \vittest }{\hbox{\bf v}}

\newcommand{\bof} {\tilde \vit }
\newcommand{\INT}{\int_0^T \int_{\Om}}

\newcommand{\E}{\varepsilon}
\newcommand{\p}{\partial}
\newcommand{\R}{\ifmmode{{\rm I} \hskip -2pt {\rm R}}
    \else{\hbox{$I\hskip -2pt R$}}\fi}
    
\newcommand{\N}{\ifmmode{{\rm I} \hskip -2pt {\rm N}}
    \else{\hbox{$I\hskip -2pt N$}}\fi}

\newcommand{\x}{{\bf x}}

\newcommand{\BEQ} {\begin{equation}}
\newcommand{\EEQ} {\end{equation}}
\newcommand{\BTHM} {\begin{Theorem}}
\newcommand{\ETHM  } {\end{Theorem}  }
\begin{document}

\begin{frontmatter}

\title{Numerical simulation of water flow around a rigid fishing net }

\author[aut1]{Roger Lewandowski}, 
\author[aut2]{G\'eraldine Pichot\corauthref{cor1}}

\address[aut1]{IRMAR, Campus Beaulieu, Universit\'e de Rennes I, 35000 RENNES, France}
\address[aut2]{IFREMER, Technop\^ole Brest Iroise, 29280 PLOUZANE , France}

\corauth[cor1]{Corresponding author. T\'el.: +33 (0)2 23 23 65 46; fax: +33 (0)2 23 23 67 90}
\ead{Geraldine.Pichot@ifremer.fr}

\begin{abstract}
This paper is devoted to the simulation of the flow around and inside a rigid axisymmetric net. We describe first how experimental data have been obtained. We show in detail the modelization. The model is based on a Reynolds Averaged Navier-Stokes turbulence model penalized by a term based on the Brinkman law. At the out-boundary of the computational box, we have used a "ghost" boundary condition. We show that the corresponding variational problem has a solution. Then the numerical scheme is given and the paper finishes with numerical simulations compared with the experimental data.
\end{abstract}


\end{frontmatter}

\section{Introduction}

Recent experimental works \cite{BCP97} show that there are less and less fish in the ocean because of intensive industrial fishing. Improvement of the selectivity of fishing nets is a major challenge to preserve fishing resources. There are still too many juvenile fish and fish with no market value are thrown overboard, leading to a real deterioration of the marine ecosystem. Solutions must be found to allow those fish to escape from the net when caught. 

\medskip
Selectivity involves a better understanding of the coupling process between the net, the surrounding flow and the fish. Measurements at sea could give some information but they are costly, difficult to perform and not easily reproducible (moving net, weather conditions, etc). Therefore, one needs to develop a numerical tool to simulate this complex mechanical system. 
\medskip

The mechanical system made of the elastic net alone in a given laminar uniform flow with very simple interaction laws has been studied already, see for instance in 
\cite{DP99}, \cite{LLPC04} and \cite{BO99}. A first approach of simulations of the flow 
around an axisymmetric rigid net has already been performed in \cite{Vincent}. 
To this point, to our knowledge, no model exists for dealing with 
the complex question due to the fish.  Finally, there is also no work concerning the coupling of an elastic net with the flow. Furthermore, it seems that today the numerical simulation of the complete system net/flow/fish does not exist. 
\medskip

In this paper we deal with the study of the flow around and inside a rigid net in the axisymmetric case. Indeed, the code written in \cite{Vincent} cannot be extended to the fully 3D case. Therefore, the coupling of the deformation of an extensible net with the fluid cannot be considered using this code. Then we have sought a mathematical model that we have  tested in the axisymmetric case and that can be extended  to the fully 3D case. We have written the corresponding numerical code and performed several simulations to fit the physical constants. Recent investigations have proved already that 3D extension is possible and is currently under progress (see \cite{GP07}). This allows to believe that it will be possible in the future to couple our fluid code to an elastic code for the net to simulate the system fluid/net.

\medskip
Our study starts from experiments performed at the IFREMER's tank of {\sl Boulogne-sur-Mer} (France). A net model rigidified by a resin (see Fig. \ref{maquette} below) was built and velocity components were measured during two experimental campaigns. The first one (see \cite {GG05}) used a Laser Doppler Velocimeter (LDV) technique to get velocity components along different profiles. The second one conducted by the second author of the present paper made use of a Particle Image Velocimeter technique (PIV). This last campaign emphasizes the locations of turbulent structures in the surrounding of the net thanks to instant pictures of the flow. It also gives a good overview of the mean flow by averages of pictures. Concerning the velocity profiles, similar shape were obtained with the two techniques, except slightly lower value with the PIV. In term of accuracy, the LDV technique is much better, that is why the LDV profiles were chosen as the reference experimental data to validate our code, for example see Fig. \ref{S18P23} to \ref{S18P67} at the end of the paper. It is striking how well the experimental velocity data fit with the numerical velocity profiles given by the code. 

\medskip

The experiments show that the flow we have to simulate is turbulent. Therefore,  one needs a turbulent model. Yet,  we have done simulations by using only the Navier-Stokes equations and we did not obtain accurate results. Therefore, we cannot bypass the Turbulent model. We have adapted to the present case a classical RANS one order turbulent closure model (see for instance \cite{LL06}, \cite{RL97B}, \cite{RL97}). It is made of an equation for the turbulent kinetic energy (TKE) and eddy viscosities functions of the TKE into the Navier-Stokes averaged equations. The mixing length has been chosen equal to the local mesh size. 

\medskip

Another important feature of the considered system is that the net behaves like a porous membrane. Taking our inspiration in \cite{GA91} combined to \cite{Angot},  
\cite{Khadra} and \cite{MI}, we have modeled the net as a porous membrane by penalizing the averaged Navier-Stokes equation with an additional linear term like in the so-called Brinkman Law. One considers the net as a fictitious domain and one solves the fluid equations in the flow domain as well as in the net domain.  However it is an open problem to validate mathematically this part of the modelization by using the homogeneization theory. 
We only notice that after a right choice of the permeabilty function $K$ (see subsection 
\ref{APT}) the model yields numerical simulations which fit very well with the experimental 
data. 

\medskip

The other last important feature of our mathematical model is the boundary conditions at the border of the computational box. On the lateral boundaries, one impose the classical no slip condition. At the incoming boundary, the flow is a given flow. The problem is what to do at the outcoming boundary. The natural and classical boundary condition should be $\boldsymbol{\sigma} . {\bf n} = 0$ where $\boldsymbol{\sigma}$ is the strain rate tensor. But as observed in \cite{FB96}, 
one risks artificial eddy reflexions. Moreover, with such a boundary condition we are not able to obtain {\sl \`a priori} estimates. To overcome this difficulty, we have adapted the ideas of \cite{FB96} to the turbulent case. To do this, we have replaced the natural condition by a so-called "ghost condition", the technical condition $(\ref{BCF})$ below. This condition becomes the natural one when the flow is laminar at the incoming and outcoming boundary (see Remark  \ref{R41}). Therefore when observing that far from the net the flow remains laminar, we can still take 
$\boldsymbol{\sigma} . {\bf n} = 0$ at the outcoming boundary. This is what we did in the numerical simulations. But we stress that the complicated condition $(\ref{BCF})$ is inescapable when dealing with the general mathematical problem. 

\medskip

Our model is given by the system $[(\ref{NS1}), ..., (\ref{NS9})]$ and the assumptions are summerized by  $[(\ref{H1}), ..., (\ref{H8})]$. For the sake of simplicity, we have chosen to study the general mathematical problem in the 2D case thankfully the axisymmetric case can be easily derived,  but technical modifications are necessary (see for instance in 
\cite{DBM99}). The existence result stated in Theorem
\ref{THM0908} is our main theoretical contribution in this paper. Uniqueness is an open problem, as well as the general 3D case. 
 
 \medskip
 
 The numerical scheme uses the finite element method in space, an implicit scheme in time for the velocity equation and a semi-implicit scheme for the equation satified by the TKE. The parameters settings are defined in section \ref{PARSET}. As shown at the end of the paper, the numerical results fit remarkably with the experimental datas. 
 
\medskip

The paper is organized as follows. We start by giving some indications on the experimental framework, then the modelization is  described followed by the mathematical analysis. The last part of the paper is devoted to the numerical simulations and the numerical results. 

\section{Experimental framework}

\begin{figure}[htbp] 
\begin{center}
\includegraphics [scale=0.4] {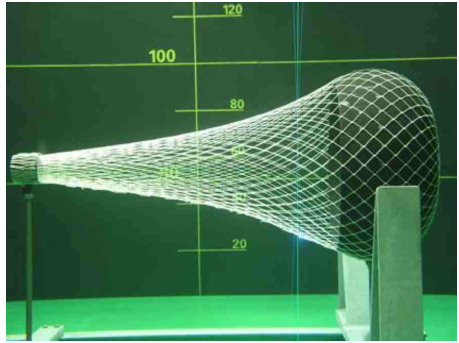} 
\caption{\footnotesize Model of cod-end net built at IFREMER - Boulogne-sur-Mer} 
\label{maquette}
\end{center}
\end{figure}

Experiments have been carried out at the IFREMER center of {\it Boulogne-sur-Mer}. Velocity profiles have been measured inside and around a rigid resin made model built by the Boulogne-Sur-Mer IFREMER team (Fig. \ref{maquette}). This model is like an axisymmetric rigid 1/6 scaled cod-end net with diamond-shaped meshes. The end of the net is  filled with a resin mass modelling a one ton catch of fish and trawled with a speed of 1.25 m/s. The net profile as well as the catch geometry have been derived from an image processing technique. 

Note that working on a rigid axisymmetric structure excludes accounting for the hydrodynamical forces exterted on the net. Moreover, it restricts the study to an axisymmetric geometry. But, at least measurements are possible and mathematical flow models can be tested.

\medskip
The model is 1 m long and has an outer maximal diameter of 0.45 m. It is maintained with a frame and set at the bottom of the IFREMER tank. This tank enables performance of flow measurements with velocities between 0.2 and 2 m/s. The estimation of the velocity to apply in the tank comes from a Froude similitude yielding an entrance velocity in the tank equal to 0.51 m/s. 

\begin{figure}[h!] 
\begin{center}
\includegraphics [scale=0.2]{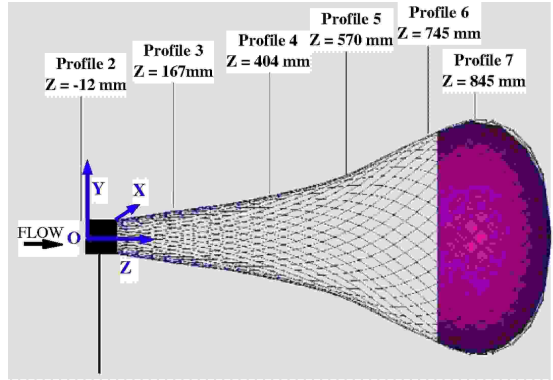} 
\caption{\footnotesize Profiles considered of the LDV measures} 
\label{profils}
\end{center}
\end{figure}

\medskip
Hydrodynamical measurements have been performed along several profiles (see Fig. \ref{profils}). 
  
\medskip
One defines a cartesian reference in the tank, the origin being set at the entrance of the net. 

\begin{figure}[htbp] 
\begin{center}
\includegraphics [scale=0.4]{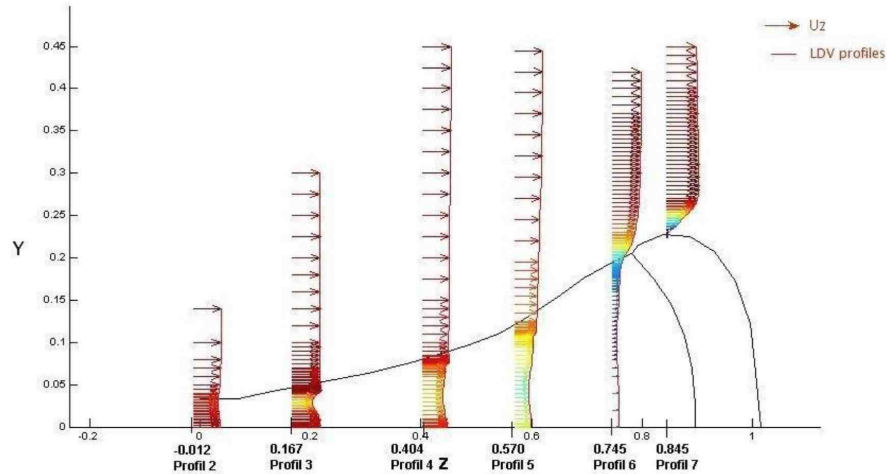} 
\caption{\footnotesize LDV profiles for the z component of the velocity.} 
\label{zcomp}
\end{center}
\end{figure}

\medskip
A Laser Doppler Velocimeter (LDV) technique was used to to collect the z and y components of the mean velocity (measures are time averaged). The z velocity component is the main one we study since it has the direction of the entrance flow, and thus the higher values (see Fig. \ref{zcomp}). 

\section{Modelization}

Our model relies on three features: 

\begin{enumerate}
\item Seeing the net, in the fluid point of view, as a porous membrane. The goal is then to define in which manner the fluid is authorized to flow through the net; 
\item Directly taking the net and the catch into account  in the averaged Navier-Stokes equations, which leads to averaged Navier-Stokes/Brinkman equations. This way, the boundary conditions at the frontiers of the obstacles are implicitely imposed;
\item Adding a one equation turbulence model to close the system.
\end{enumerate}


Our study deals with the mean flow. One can make the assumption that the mean flow around the net is axisymmetric. 

\subsection{Axisymmetric hypothesis} \label{GEOM}
\begin{figure}[htbp] 
\begin{center}
\includegraphics [scale=0.6]{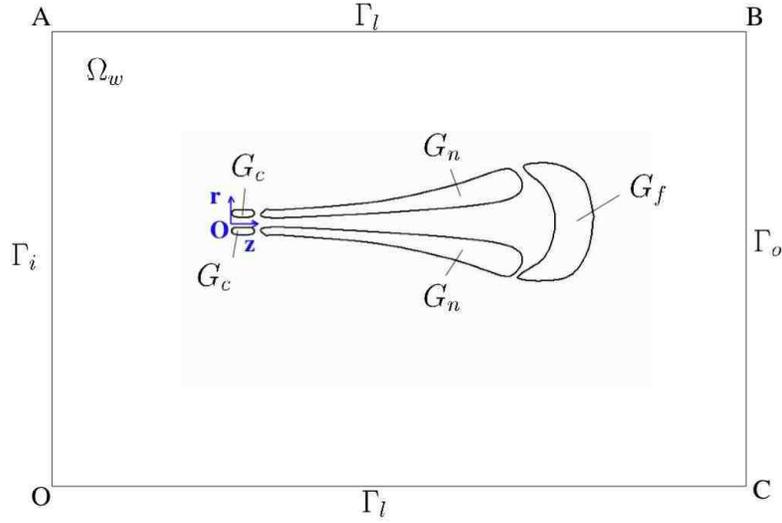} 
\caption{\footnotesize Geometry and notations} 
\label{geometry}
\end{center}
\end{figure}

Assume the cod-end net is embedded in a cylinder full of water. Let us consider an axisymmetric deformation of the net (See Fig. \ref{geometry}). As the net is modeled by a porous membrane the problem reduces to a 2D one, provided an axisymmetric hypothesis of the flow. We admit this hypothesis is a strong one but reasonable in the case of the study of the mean flow, since turbulent structures are smoothed by the averaging.

 \medskip
\medskip
In the following, one notes 
\begin{itemize} 
\item $\Omega_w$ the domain occupied by the water,
\item $G_{n}$ the net domain,
\item $G_{f}$ the the fish domain, 
\item $G_{c}$ the domain formed by the frame at the entrance of the net model,
\end{itemize} 
$$ \Omega = \Omega_{w} \cup G , \quad G=G_{n} \cup G_f \cup G_{c}.$$

Using the assumption of an axisymmetric flow and the model of an axisymmetric equivalent membrane to describe the net, cylindrical coordinates $(O,r,z,\theta)$ are used in the simulations. At a fixed value of $\theta$, the mathematical problem reduces to a 2D one. The artificial cylinder reduces to a rectangle in the reference $(O, r, z)$ and the sides of this rectangle are called $\Gamma_i$, $\Gamma_l$ and $\Gamma_o$ (see Fig. \ref{geometry}).

\subsection{A membrane model for the net}
Finite elements and finite volumes methods are known to be the common numerical methods to compute fluid dynamics. A mesh is built to discretize the fluid domain. The difficulty of the netting is that it is composed of a great number of meshes. Generating a body-fitted fluid mesh, that is a mesh lying on the nodes and the twines of the net, would be far too complex and computer time consuming. Then, an exact description of the net would be too demanding in computer resources to be conceivable. Another model has to be found.

\medskip
In the literature, one finds a model of an axisymmetric membrane to deal with an axisymmetric porous structure immersed in a fluid (see \cite{Vincent}). 

\medskip
In \cite{Vincent}, the equations are set on the structure location to express a mass transfer in the normal direction to the structure and slip effects in the tangential direction. Then, the tangential velocity, denoted $u_t$, is set to be governed by Shaffman's law and the normal velocity, denoted $u_n$, by Darcy's law.

\medskip
This leads to express the velocity components at the wall of the axisymmetric structure by:

\begin{equation}
\left \{
\label{Shaffman}
\begin {array}{lll}
\displaystyle u_t & = & \displaystyle B \frac {\partial u_t}{\partial n},\\\\
\displaystyle u_n & = & - K \nabla p, 
\end{array}
\right.
\end{equation}

where $n$ is the outer normal of the structure, $p$ the fluid pressure, $K$ a permeability tensor found experimentally, and $B$ a coefficient dependent on the tangential velocity and then deduced from numerical experiences.

\medskip
To solve the problem,  one builts a cartesian mesh from the geometry of the membrane, using curvilinear coordinates. The velocity and pressure unknowns are computed using a finite differences method. 

\medskip
A drawback of  this method is that it is based on a cartesian mesh which is not convenient to build and to refine locally in the case of a complex net profile. This work then cannot easily be generalized to the case of a 3D deformation of the net. One has to find a flow model that allows a future coupling with a moving net.

\medskip
Let us keep the idea of seeing the net as a porous medium, as this assumption has the advantage of making the numerical programming simpler insofar as twines and nodes are no longer taken into account. Then, consider the net and the catch as domains with a given permeability. 
\medskip

As shown in Fig. \ref{geometry}, the domain $G_n$ delimiting the net has a thickness much larger than the diameter of net twines (which is typically 3 mm). This idea actually came from the analysis of the velocity profiles in the $z$ direction obtained by the LDV measurements (see Fig. \ref{zcomp}).

\medskip

One notices on the LDV profiles (see Fig. \ref{zcomp}) inner minima of the z velocity component. See Fig. \ref{S18P23}-\ref{S18P45}-\ref{S18P67} in the following section for a zoom of each profile. Those minima have been noted down (see Table \ref{Tabmin}). The inner profile of the membrane has been drawn thanks to those values. The outer profile is in agreement with the profile of the model.

\begin{table}[!h] 
\centering \includegraphics [scale=0.8]{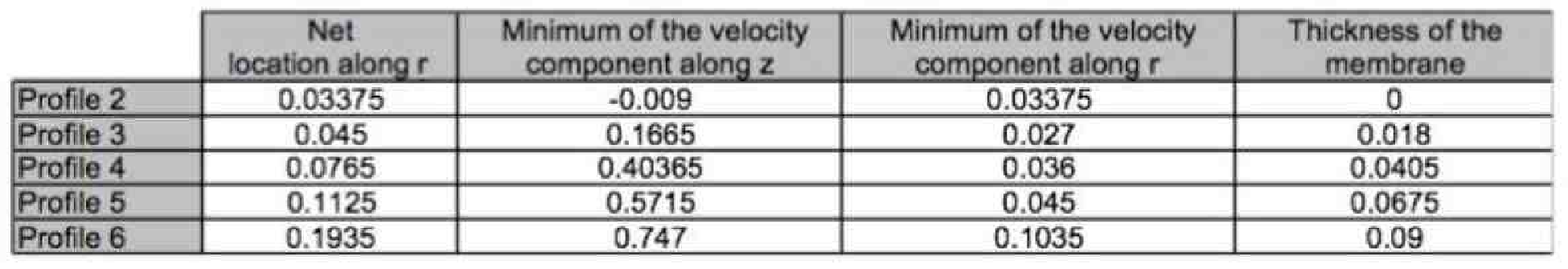}  \caption{\footnotesize}
\label{Tabmin}
\end{table}  

\medskip
This approach avoids a costly mesh generation. However, it comes with the difficulty of determining which permeability to apply in the different media. The next part is devoted to explain how those media are taken into account directly in the equations.

\subsection{\label{APT} A penalization technique}
The second feature of our model relies on a penalization method that allows us to take the presence of the obstacles into account directly in the fluid equations \cite{Khadra}, \cite{MI}, \cite{Angot}. The method consists in solving "fluid" equations in the entire domain, even in the net and catch domains. The net domain is seen as a porous medium, and the catch domain as a solid medium, where a no-slip boundary condition should hold. Those media are explicitly included in the fluid equations by the addition of a penality term of the velocity, namely $\displaystyle \frac{1}{ K({\bf x})} \, {\bf u}$. This leads to  Navier-Stokes/Brinkman equations. Notice that such laws have been derived from an homogeneization process in other situations, as in \cite{GA91}. This theoretical question remains open in this particular context. The function $K({\bf x})$ varies from one domain to another. It is a permeability parameter that is very small in the solid domains, e.g. the catch, to force the velocity to be zero, and very high in the fluid domain, so that the averaged Navier-Stokes equations hold and are set to a defined value or function in the porous domain (here in the net domain) depending on its permeability. 

\begin{figure}[!h] 
\begin{center}
\includegraphics [scale=0.9]{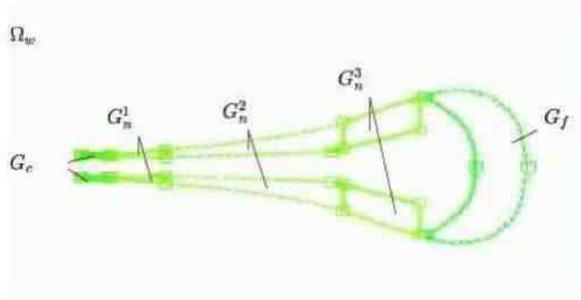} 
\caption{\footnotesize Decomposition of the net domain - Notations} 
\label{zoomgeometrie}
\end{center}
\end{figure}

\medskip
At a first glance, the function $K({\bf x})$ is set to be constant by parts. The net domain is decomposed in three parts, $G_n^i$, i=1, 2, 3 (see Fig. \ref{zoomgeometrie}) of constant permeability that is all the more important as we are closer to the catch (see Part \ref{simu}). In a future work, we will try to make it depend on the mesh opening, the mesh angle between the mesh and the local flow.  

\subsection{Addition of a turbulence model}
The third point comes with the average of the Navier-Stokes/Brinkman equation, since Direct Numerical Simulation would not be able to treat a problem with such a high Reynolds number (here $Re \cong 10^5$, using as reference length the maximal diameter of the catch, i.e. 0.45 m, and the entrance velocity as reference velocity that is equal to 0.51 m/s). A kind of Reynolds Averaged Navier-Stokes (RANS) turbulence model is then added to close the system of equations. It consists of one equation for the turbulent kinetic energy. The averaged NS/Brinkman equation and the turbulent kinetic energy equation are coupled by the means of a eddy viscosity, denoted $\nu_t$.

\section{Description of the mathematical problem}
 
\subsection{The domain} 
 We return back to the description of the geometry. As already said, the flow under study is axisymmetric. In order to avoid technical complications, we have chosen to study the mathematical problem set in a domain in $\R^2$. 
 We refer to \cite{DBM99} to go in further developments in the axisymmetric case. 
 \medskip

\begin{figure}[htbp] 
 \begin{center} 
 {\includegraphics [scale=0.50]{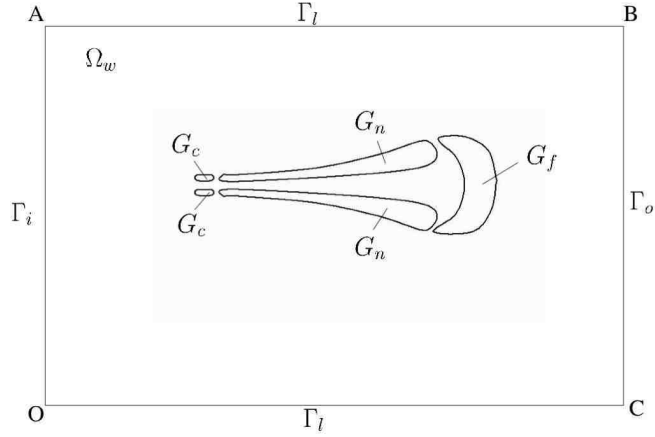}} 
\caption{\footnotesize Description of the domain } 
 \end{center}  
 \end{figure}

 The boundary $\Gamma$ of the computational box is defined by the input board $\G_i$, the lateral boards $\G_l$ and the artificial output board $\G_o$,  
  \begin{equation} \label {Board} \begin{array} {ll}   \G_i = [O, A], & 0= (0,0), \quad A=(0, \alpha), \\
 \G_l = [C, O] \cup [A, B], & B=(\beta, \alpha), C = (\beta, 0), \\
 \G_o = [B, C] , & ~Ê\\
 \Gamma = \G_i \cup \G_l \cup \G_o. & ~
 \end{array} \end{equation}

 \subsection{The equations} 
  The unknowns are : 
 \begin{itemize} 
 \item the mean velocity vector field $\vit = \vit (t, \x) = (u^1 (t, \x), u^2 (t, \x))$, 
 $\x = (x, y)$,
 \item the mean pressure scalar field $p = p (t, \x)$,
 \item the turbulent kinetic energy $k = k(t, \x)$
  \end{itemize}
  One defines the deformation tensor $\boldsymbol {\E}$ 
  by 
  \begin{equation}\label {DEF}
   \boldsymbol {\E}(\vit) = {\g \vit + \g \vit^T \over 2}. \end{equation} 
   The turbulent strain rate stress tensor $\boldsymbol {\sigma}$ is defined by 
   \begin{equation} \label{Tens}  
   \boldsymbol {\sigma} (\vit, p, k) =  2 \,\nu_t (k,\x) \boldsymbol {\E}(\vit) - p \, {\bf Id}.\end{equation} 
The Reynolds Averaged Navier-Stokes turbulent closure model of order one including the Brinkman laws, 
is given in $[0, T] \times \Omega$ ($T>0$) by the following equations, where $\E >0$ is fixed, 
 
\begin{equation} \label {NS} \begin{array} {l} \displaystyle
   \p_t \vit + (\vit \nabla ) \vit - \nabla \cdot  \boldsymbol {\sigma} (\vit, p, k) + 
  \left ( {1 \over \E}( \Ind_{G_f \cup G_c}  ) + {1 \over K(\x)}   \Ind_{G_n} + \E \Ind_{\Omega_w} \right ) \vit
   = {\bf 0},  \\ 
\nabla \cdot \vit = 0,   \\
\p_t k + \vit \, . \g k - \nabla \cdot (\mu_t (k,\x) \nabla  k  ) = 2 \nu_t (k,\x) |\boldsymbol {\E}(\vit) |^2 - {\cal E} (k, \x). 
\end{array}
  \end{equation}
In the equations above, $\nu_t$ and $\mu_t$ are the eddy viscosities and 
${\cal E}$ the backward term. Their analytical expressions are given in section \ref{eddy} below. 
  
  \subsection{The boundary conditions and the initial data} 
  \subsubsection{Boundary conditions} 
  The input field $\vit_{\hbox{\tiny I}} = (u_{\hbox{\tiny I}}, 0)$ at the boundary $\G_i$ is a data of our problem. The boundary conditions we consider are the following.

  \begin{eqnarray}  \label {BC}     && \hbox{on } \G_i : \quad  \vit = \vit_{\hbox{\tiny I}} = (u_{\hbox{\tiny I}}, 0) , \quad \quad k = 0,  \\
 && \hbox{on } \G_l : 
\quad  \vit = {\bf 0}, \quad  \quad k = 0,  \\
&& \label{BCF} \hbox{on } \G_o : 
\quad  \left \{ \begin{array}{l} \displaystyle \boldsymbol {\sigma}(\vit, p, k). {\bf n} 
=  -{1 \over 2} (\vit .  {\bf n})^- (\vit-\vit_{\hbox{\tiny I}}) 
+ (\vit .  {\bf n}) \vit_{\hbox{\tiny I}}
\\ \displaystyle
k=0.   \end{array} \right. 
\end{eqnarray}

In the formulae above, $\vit_{\hbox{\tiny I}}$ denotes the field equal to 
$ (u_{\hbox{\tiny I}} (x-\beta, 0))$ on $\G_o$.  One uses the boundary condition $(\ref{BCF})$ in order to avoid eddy reflections at the open boundary $\G_o$ and to be able to prove the existence of a dissipative solution to the system $(\ref{NS})$.

\begin{Remark}\label{R41} The natural boundary condition for the velocity at $\G_o$ should be \\ ${\sigma}(\vit, p, k). {\bf n} = {\bf 0}$. In 
\cite{FB96}, the authors study the case of the Navier-Stokes equations without a turbulence model 
and in a channel without a rigid body. They remark that the boundary condition  ${\sigma}. {\bf n} = {\bf 0}$ yields numerical eddy reflexions at the out open boundary. Moreover, the existence of a dissipative solution is not known in such a case because of a term 
$ \int_{\G_0} (\vit . {\bf n} ) | \vit |^2$
which appears in the energy equality due to the convection. Without additional information on the sign of $(\vit . {\bf n} )$ at $\G_0$, no 
{\sl \`a priori }Êestimate is avaible. 
This is why the authors in 
\cite{FB96} change the boundary conditions. We also change  the boundary conditions by an adpatation to the case of our turbulence model. Notice that when the flow is laminar at $\G_0$ and  $(\vit . {\bf n} ) >0$ on $\G_0$, the boundary condition reduces to the classical one up to the term 
$(\vit . {\bf n} )\vit_{\hbox{\tiny I}}$. This is an additional forcing term. Without this term, it is easy checked that one can only derive an {\sl \`a priori} estimate when a smallness assumption on $\vit_{\hbox{\tiny I}}$ is satisfied, an assumption which would restrict the problem to a laminar one. Therefore, this term seems to look coherent when the flow  is turbulent at the incoming boundary. However, in the numerical simulations we have taken ${\sigma}(\vit, p, k). {\bf n} = {\bf 0}$. Indeed, the experiments suggest that the flow is laminar far from the net. Therefore, our choices are in concordance with reality and yields a rigorous mathematical analysis. 
\end{Remark} 

\begin{Remark} For convenience and the sake of simplicity, we have chosen to develop the theoretical part by fixing $k=0$ at $\G_o$. A more natural boundary condition at $\G_o$ is 
$\mu_t {\p k \over \p {\bf n}}   = 0$. This is the condition that we use in the numerical simulations. From the mathematical viewpoint, we then have to impose
$\mu_t {\p k \over \p {\bf n}}  = - (\vit.{\bf n})^{-} k$ at $\G_o$. Therefore the discussion in remark \ref{R41} above holds in this case. However, this boundary condition yields serious mathematical complications that would have been out of the scope of this paper. In subsection \ref{NTCTKE} we give some explanations  about this case.  
\end{Remark} 

\medskip
Throughout the paper, we assume that 
\begin{equation} \label{H44} u_{\hbox{\tiny I}}\in H^{3/2}_{00} (\G_i). 
\end{equation}
 
 \subsubsection{Initial data} 
  The initial data are specified by 
 \begin{eqnarray} && \forall \, \x \in \Om, \quad  \vit (0, \x) = \vit_0 (\x) \in (L^2(\Om))^2, \\
&&   \forall \, \x \in \Om, \quad   k(0, \x) = k_0 (\x) \in L^1 (\Om).  \end{eqnarray}
Moreover, we shall assume that $\vit_0$ satisfies the compatibility conditions
\begin{eqnarray} && \label {H11} \g \cdot \vit_0 = 0, \\
&&  \label{H22}  \vit_0 . {\bf n} = u_{\hbox{\tiny I}}  \quad \hbox{on } \G_i,  \\
&& \label {H33} \vit_0 . {\bf n} = 0 \quad \hbox{on } \G_l.
\end{eqnarray}

  \begin{Remark} The assumption $\vit_0 \in (L^2(\Om))^2$ gives a sense to 
  $\vit_0 . {\bf n} $ in the spaces $(H^{3/2}_{00} (\G_i))'$ and 
  $(H^{3/2}_{00} (\G_l))'$, making $(\ref{H22})$ and $(\ref{H33})$ consistant  as a consequence of $(\ref{H44})$. 
   \end{Remark}

 \subsection{\label{eddy} The eddy viscosities and main terms} 
 \subsubsection{Eddy viscosities} 
 The eddy viscosity function $\nu_t$ is a $C^1$ non negative bounded function of $k$ and $\x$ equal to $\nu_0 + \ell (\x) \sqrt {\tau + |k|}$ when $ |k | \in [0, k_c]$ for a given $k_c$ and $\rho >0$ is fixed. The viscosity $\nu_t$ is thus given by 

 \begin{eqnarray} \label{nut} 
 && \nu_t (k,\x)  =  \nu_0 + \ell (\x) \sqrt {\tau + |k| } , \quad \hbox{when } |k| \le k_c, \\
&& \nu_t (k,\x )  =  v_2 ,  \quad  \hbox{when } |k| \ge k_c +1,
\end{eqnarray} 
\begin{equation} \left \{  \begin{array} {l}\nu_t (k,\x )  =  \displaystyle \left ( \frac {l(x)}{2\sqrt{\tau + k_c}} +2 v_1 - 2 v_2 \right ) k^3  
 \displaystyle +   \displaystyle \left ( \frac{l(x)(-3 k_c-2)}{2\sqrt{\tau + k_c}} + \right. \\ \left. (v_1 - v_2)(-6 k_c-3) \right ) k^2 
\displaystyle + \displaystyle \left ( \frac{l(x)(1+3 k_c^2 + 4 k_c)}{2\sqrt{\tau + k_c}} + (v_1 - v_2)(6 k_c^2 + 6 k_c) \right ) k 
\displaystyle \, + \\  \displaystyle v_1 + \frac{l(x)(-k_c^3 - 2 k_c^2 - k_c)}{ 2\sqrt{\tau + k_c}} 
 \displaystyle + (v_1 - v_2)(-2 k_c^3 - 3 k_c^2) \quad  \hbox{when }  k_c < k < k_c +1. 
\end{array} \right. \end{equation} 
 where 
 $  \tau > 0, \quad  k_c >0, \quad 
  v_2 > v_1 = \nu_0 + \ell (\x) \sqrt {\tau + k_c }. $
  
  \begin{figure}[htbp] 
  \begin{center} 
  {\includegraphics [scale=0.4]{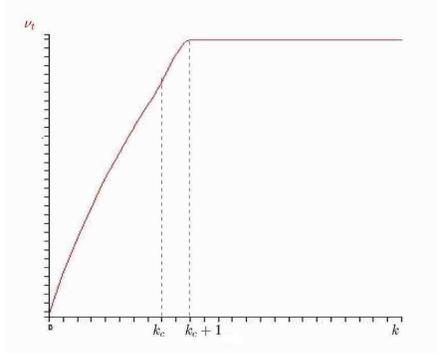}} 
  \caption{\footnotesize Shape of $\nu_t$}  
  \end{center}   
  \end{figure}

 The function $\ell (\x)$ is a local scale of the flow. It is a non negative bounded $C^1$ function of $\x$ on $\Om$ with
 \begin{equation}  \label {IN1} \forall \, \x \in \Om, \quad 0 < \ell_0 \le   \ell (x) \le L_0 < \infty.\end{equation} 
 The eddy diffusivity $\mu_t$ is of the same form as $\nu_t$ and 
\begin{equation}\label {VISC3} \nu_0 + C \ell(\x) \sqrt {\tilde \tau + | k | } \quad \hbox{on the range} \quad  [0, k_c], \end{equation} 
for $C>0$ and $\tilde \tau>0$
 fixed coefficients. 
 
 \subsubsection{Backward term} 
 The backward term ${\cal E} (k,\x)$ is given by the formula 
 \begin{equation} \label{Back1} 
{\cal E} (k,\x) = {1 \over \ell (\x) } k \sqrt k. \end{equation}

\subsubsection{Permeability}Ê
The permeability function $K(\x)$ Êis a continuous function that satisfies 
 \begin{equation}  \label {IN2} \forall \, \x \in \Om, \quad 0 < K_0 \le   K (\x) \le K_1 < \infty.\end{equation} 
 In the remainder, one shall set 
 \begin{equation} {\cal P} (\vit) (t, \x) =   
 \left ( {1 \over \E}( \Ind_{G_f \cup G_c} (\x)  ) + {1 \over K(\x)}   
 \Ind_{G_n} (\x) +  \E  \Ind_{\Omega_w} \right ) \vit (t, \x), 
 \end{equation}  
where $\E >0$ is fixed. 
 
 \section{\label{MATHS} Mathematical analysis} 
  \subsection{Main result}Ê 
We summerize the hypotheses: 
\begin{eqnarray} && \label {H1} \nu_t \in C^1 , \quad \forall \, (k, \x) \in \R\times \Om, \quad 
 0< \nu_0 \le \nu_t(k, \x) \le N < \infty, \\
 &&  \label {H2} \mu_t \in C^1 , \quad \forall \, (k, \x) \in \R\times \Om, \quad 
 0< \mu_0 \le \nu_t(k, \x) \le  M < \infty, \\
 && \label {H3}\ell \in L^\infty , \quad \forall \, \x \in \Om, \quad 0 < \ell_0 \le   \ell (x) \le L_0 < \infty, \\
 && \label {H4}{\cal E} (k, \x) = {1 \over \ell (\x) } k \sqrt {|k|}, \\
 && \label {H5} K \in C^1, \quad \forall \, \x \in \Om, \quad 0 < K_0 \le   K (\x) \le K_1 < \infty. \\
 && \label {H6} \vit_0 \in L^2(\Om), \quad  \g \cdot \vit_0 = 0, \quad \vit_0 . {\bf n}\vert_{\G_i}  = u_{\hbox{\tiny I}}, \quad 
\vit_0 . {\bf n}\vert_{\G_l} = 0, \\
&& \label {H7} k_0  \in L^1 (\Om), \quad k_0 \ge 0 \, a.e, \\
&& \label {H8} \vit_{\hbox{\tiny I}} \in H^{3/2}_{00} (\G_i).
 \end{eqnarray}
 The problem is the following 
\begin{eqnarray} && \label {NS1} \displaystyle
   \p_t \vit + (\vit \nabla ) \vit - \nabla \cdot  \boldsymbol {\sigma} (\vit, p, k) + 
 {\cal P}(\vit)
   = {\bf 0},  \\ 
&& \label {NS2}  \nabla \cdot \vit = 0,   \\
&& \label {NS3} \p_t k + \vit \, . \g k - \nabla \cdot (\mu_t (k,\x) \nabla  k  ) = 2 \nu_t (k,\x) |\boldsymbol {\E}(\vit) |^2 - {\cal E} (k, \x). \\
&& \label {NS4} \forall \, \x \in \Om, \quad  \vit (0, \x) = \vit_0 (\x), 
\\
&&  \label {NS5}   \forall \, \x \in \Om, \quad   k(0, \x) = k_0 (\x), \\
&&  \label {NS6}  
\vit \vert_{\G_i} = \vit_{\hbox{\tiny I}} = (u_{\hbox{\tiny I}}, 0), 
\quad k\vert_{\G_i } = 0, \\
&&  \label {NS7} \vit \vert_{\G_l} = {\bf 0}, \quad k\vert_{\G_l} = 0, \\
&&  \label {NS8} \boldsymbol {\sigma}(\vit, p, k). \, {\bf n} \vert_{\G_o} 
=  -{1 \over 2} (\vit .  {\bf n})^- (\vit-\vit_{\hbox{\tiny I}}) 
+ (\vit .  {\bf n}) \vit_{\hbox{\tiny I}}, \\
&&  \label {NS9} \displaystyle
k  \vert_{\G_o} = 0. 
\end{eqnarray}
Our main result is the following. 
 \begin{Theorem} \label{THM0908} Assume that hypotheses $[(\ref{H1}) ... (\ref{H8})]$ hold. Then Problem 
 $[(\ref{NS1})... (\ref{NS9})]$ admits a solution $(\vit, p, k)$ on any time interval $[0, T]$  in the sense of the distributions, where 
 \begin{eqnarray} && \label {REG1} \vit \in L^2([0,T], (H^1 (\Om))^2 ) \cap L^\infty([0,T], L^2(\Om)), \\
 && \label {REG2} p \in L^2 ([0,T] \times \Om),\\
 && \label {REG3} k \in L^\infty([0,T], L^1(\Om)) \cap  ( \bigcap_{p<{4 \over 3}}L^{p} ([0,T], W^{1, p} (\Om) )  )   .
 \end{eqnarray} 
 \end{Theorem}

\begin{Remark} Uniqueness remains an open problem. 
\end{Remark} 
  
  \subsection{Lifting the boundary condition} \label {lift}
   \subsubsection{Auxiliary Stokes Problem} 
   In this section, we describe how to lift the boundary conditions to reduce the problem to a problem with homogeneous boundary conditions on $\G_i \cup \G_l$, as it is usually done in mathematical problems where Navier-Stokes Equations are involved. 
  \medskip
  
 Recall that $\Om_w$ is the water domain and $G$ the net domain (see section \ref{GEOM}).  
 \medskip
 
The incoming flow $  \vit_{\hbox{\tiny I}} $ is prescribed at the boundary 
$\G_i$. We define 
$  \vit_{\hbox{\tiny I}} $ on the output boundary $\G_o$ and still denote it 
 by $\vit_{\hbox{\tiny I}}$, the field defined by 
$$ \forall \, \x = (x, y) \in \G_o, \quad  
\vit_{\hbox{\tiny I}} (x,y) = (u_{\hbox{\tiny I}} (x-\beta, 0)).$$
Let us consider the Stokes problem 
   \begin{equation} \label {ST} \begin{array} {l} 
 - \Delta \vittest_0 + \g q_0 = {\bf 0} \quad \hbox{in }Ê\Om_w, \\
 \hskip 0.5cm \g \cdot \vittest_0 = 0 \quad \hbox{in }Ê\Om_w, \\
  \hskip 0.5cm \vittest_0 = {\bf g} \quad \hbox{on }Ê\G \cup \p G,
  \end{array}
  \end{equation}
  where $Ê\Om_w$ is the water domain, $G$ the domain delimited by the net and 
  ${\bf g}$ is the field defined by 
  \begin{equation} \label {BC2} \begin{array} {ll}   
   \hbox{on } \G_i \cup \G_o,  &  {\bf g} = \vit_{\hbox{\tiny I}},   \\
 \hbox{on } \G_i \cup \p G,  
 &  {\bf g} = {\bf 0},   \end{array} 
  \end{equation} 
Notice that the following compatibility condition is satisfied: 
\begin{equation}\label{Comp}  \int_\G {\bf g} . {\bf n} = {\bf 0}.\end{equation}
In the following, we note 
$$ L^2_0(\Om) = \{ q \in L^2(\Om); \, \int_\Om q(\x) \, d\x = 0 \} .$$
  
\begin{Theorem} \label{ST1} Assume that  $u_{\hbox{\tiny I}} \in H^{3/2}_{00} (\G_i)$ (assumption $(\ref{H8})$). Then Problem $[(\ref{ST})-  (\ref{BC2})]$ has a unique solution 
  $(\vittest_0, q_0) \in H^2 (\Om_w) \times (H^1 (\Om_w) \cap L^2_0 (\Om_w))$. 
  \end{Theorem} 
  
  {\bf Proof}. On one hand, it is established by Corollary 5.9 in \cite{DBM03}  that 
  ${\bf g} \in [H^{3/2} (\G)]^2$ because $u_{\hbox{\tiny I}} \in H^{3/2}_{00} (\G_i)$. 
On the other hand, ${\bf g}$ satisfies the compatibility condition $(\ref{Comp})$. Moreover, $\Om_w$ is a convex polygon in dimension 2. Therefore, applying  Theorem 5.4 and Remark 5.6 in \cite{GR86} \S 5   (see also \cite{PG78}), one knows the existence of a unique $(\vittest_0, q_0) \in H^2 (\Om_w) \times (H^1 (\Om_w) \cap L^2_0 (\Om_w))$ solution to Problem $[(\ref{ST})- (\ref{BC2})]$.

   \begin{Remark} In practical situations,  $\vit_{\hbox{\tiny I}}$ is
 a Poiseuille flow. Therefore, one has 
 $$ u_{\hbox{\tiny I}} (x,y)=   u_{\hbox{\tiny I}}(y) =  D y (\alpha - y ),$$
 where $D$ is a constant. We first note that $u_{\hbox{\tiny I}}$ is $C^\infty$ on 
 $\G_i$. Moreover, 
 one clearly has 
 $$ \int_0^\alpha  { |u_{\hbox{\tiny I}} (y) |^2 \over y} dy < + \infty \quad 
 \hbox{and} \quad  \int_0^\alpha  { |u_{\hbox{\tiny I}} (y) |^2 \over (\alpha - y)} dy < + \infty.$$
 Therefore, thanks to the definition of $H^{1/2}_{00}$ (see in \cite{LM68}, chapter 1, \S 11 or in \cite{DBM03} chapter 6), $u_{\hbox{\tiny I}} \in H^{1/2}_{00} (\G_i)$. Unfortunaly, $u_{\hbox{\tiny I}} \notin H^{3/2}_{00} (\G_i)$. Therefore, one cannot guaranty that ${\bf g} \in H^{3/2} (\G)$ by using the results above mentioned and only 
 ${\bf g} \in H^{1/2} (\G)$. In such a case, only
 $H^1$ regularity for the velocity can be obtained {\sl \`a priori} and that is not enough regularity for what follows, as we shall see in the remainder.  \end{Remark} 
 
 \begin{Remark} Since   ${\bf g} \in [H^{3/2} (\G)]^2$, the trace on $\G_o$ of $\E (\vittest_0)$ is in $[H^{1/2} (\G_o)]^4$ as well as the trace of $q_0$ on $\G_o$ is   in 
$H^{1/2} (\G_o)$. Then, because $\nu_t$ is a bounded function, for every 
$k \in L^1 (\Om)$, 
\begin{equation}\label{UR1}  \boldsymbol{\sigma} (\vittest_0, q_0, k) \in [H^{1/2}(\G_o)]^4. 
\end{equation} 
\end{Remark}

From now, one still denotes by $\vittest_0$ the field defined on whole $\Om$ and equal to 
$\vittest_0$ in $\Om_w$ , the 
velocity part in the solution to Problem $[(\ref{ST})-  (\ref{BC2})]$, and equal to $0$ inside $G$. Since \begin{itemize}
\item $H^2 (\Om_w) \subset L^\infty (\Om_w)$
\item $\p G$ is of class $C^1$, therefore one can use Proposition IX.18 in 
\cite{HB93}, 
\end{itemize} 
one has
\begin{equation} \label{REG612}  \vittest_0 \in H^1( \Om) \cap L^\infty (\Om)\end{equation}
and 
\begin{equation} \label{110562} || \vittest_0 ||_{H^1 (\Om)} + || \vittest_0 ||_{L^\infty (\Om)} 
\le C || u_{\hbox{\tiny I}} ||_{H^{3/2}_{00} (\G_i)},\end{equation} 
where $C$ only depends on $\alpha$ and $\beta$. By extending $q_0$ by zero outside 
$\Om_w$ and still denoting the expension by $q_0$, one has 
\begin{equation}  \boldsymbol{\sigma} (\vittest_0, q_0, k) \in [L^2(\Om)]^4. 
\end{equation} 
 Notice also that  
\begin{equation}
{\cal P}(\vittest_0) = {\bf 0}. 
\end{equation}

\subsubsection{Change of variable} 
We set:
\begin{equation}
\label{uptild}
\displaystyle \vit = \tilde \vit + \vittest_0, \quad
\displaystyle p = \tilde{p} + q_0.
\end{equation}

It is straightforward to prove that  $(\tilde \vit, \tilde{p}, k)$ is 
governed by the following system:

\begin {eqnarray}
 && \label {NSH1} \left \{ \begin{array} {l} \displaystyle
   \p_t \tilde \vit + (\tilde \vit \nabla ) \tilde \vit - \nabla \cdot  \boldsymbol {\sigma} (\tilde \vit, \tilde{p}, k) \, + {\cal P} (\tilde \vit) + \\
   (\tilde \vit  \g) \vittest_0 + (\vittest_0 \g ) \bof
   + (\vittest_0 \g) \vittest_0  - \nabla \cdot  \boldsymbol {\sigma} (\vittest_0, q_0, k)
 \displaystyle 
   = {\bf 0},  \end{array} \right.  \\ 
&& \label {NSH2} \nabla \cdot \tilde \vit = 0,   \\
&& \label {NSH3}\left \{ \begin{array} {l} \displaystyle  \p_t k + \bof \, . \g k - \nabla \cdot (\mu_t (k,\x) \nabla  k  )  = 
2 \nu_t (k,\x) |\boldsymbol {\E}(\bof) |^2 - {\cal E} (k, \x) + \\
4 \nu_t (k, \x)\boldsymbol {\E}(\bof). \boldsymbol {\E}(\vittest_0) + 
2 \nu_t (k,\x) |\boldsymbol {\E}(\vittest_0) |^2- {\bf v}_0 \g k . \end{array} \right. \\
&& \label {NSH4}\bof \vert_{ t=0} = \vit_0 - \vittest_0, \quad k \vert _{t=0} = k_0, \\
&& \label {NSH5}\bof \vert_{ \G_i \cup \G_l} = {\bf 0}, \quad k \vert_{ \G_i \cup \G_l} =0, \\
&& \label {NSH6} \left \{ \begin{array} {l} \displaystyle  \boldsymbol {\sigma } (\bof, \tilde p, k) . {\bf n} \vert_{\G_o} =  \\ \displaystyle 
-{1 \over 2} 
[(\bof + \vittest_0). {\bf n}]^- \bof  \, + Ê
[(\bof + \vittest_0). {\bf n}] \vittest_0 - \boldsymbol {\sigma } (\vittest_0, q_0, k)\vert _{\G_o}  . {\bf n},   \end{array} \right. \\
&& \label {NSH7} k  \vert_{\G_o} = 0. 
  \end{eqnarray}

\subsection{Variational formulation}
\subsubsection{Functions space}
The natural space for studying Problem $[(\ref{NSH1}) - - (\ref{NSH6})]$ is the space 
\begin{equation} V = \left \{ \vittest \in (H^1 (\Om))^2; \quad \g \cdot \vittest = 0; \quad \vittest \vert _{\G_i \cup \G_l} = {\bf 0}.
\right \}Ê\end{equation} 
In order to use De Rham Theorem and have an Inf-Sup condition on the pressure, we must check that smooth vector fields with null divergence and equal to zero on $\G_i \cup \G_l$ consitutes a dense space in  
$V$. This is the goal of what follows. 

\medskip

Let $\tilde B = (2\beta, \alpha)$, $\tilde C = (2 \beta, 0)$ and let $\tilde \Om$ be the square in $\R^2$ bounded 
by the points $O$, $A$, $\tilde B$ and $\tilde C$. Let $s$ be the symmetry through the 
axis $x= \beta$, that is $s (x,y) = (2 \beta -x, y)$. 
\medskip

We also denote by $\Om^s$ 
the square bounded by the points $C$, $\tilde C$, $\tilde B$ and $B$,  also defined by 
$\Om^s = s (\Om)$. 
\medskip

Let $\tilde V$ be the set 
$$ \tilde V = \left \{ \vittest \in (H^1 (\tilde \Om))^2; \quad \g \cdot \vittest = 0; \quad \vittest \vert _{\p \tilde \Om} = {\bf 0}
\right \}$$
as well as 
$$ \tilde {\cal V} = \left \{ \vittest \in ({\cal D} (\tilde \Om))^2; \quad \g \cdot \vittest = 0 \right \}.$$
Being given ${\bf v}Ê\in \tilde V$, let ${\bf v}_r$ be its restriction to the square 
$\Om$. One obviously has ${\bf v}_r \in V$. 
\medskip

Being given ${\bf v}Ê\in V$, let ${\bf v}^e$ be its extension to $\tilde \Om$ defined as follows: 
\begin{equation} \label{ZB1}Ê\begin{array} {ll}  \forall \, \x \in \Om, & {\bf v}^e (\x) = {\bf v}Ê (\x), \\
 \forall \, \x \in \Om^s, & {\bf v}^e (\x) = {\bf v}Ê (s(\x)).\end{array} 
\end{equation} 

Notice that ${\bf v}^e \in \tilde V$ and one has 
\begin{equation} \label {ZB2}  \int_{\Om^s} | \g {\bf v}^e |^2 = 
2 \int_\Om | \g {\bf v} |^2, \quad \forall \, p \in [1, \infty[, \quad 
 \int_{\Om^s} |{\bf v}^e |^p = 2 \int_{\Om} |{\bf v} |^p. \end{equation}

Finally let ${\cal V}$ be the space made of the restrictions to $\Om$ of fields in 
$\tilde {\cal V}$, which means 
\begin{equation} {\cal V} = \left \{ {\bf v}Ê\in [C^\infty (\Om)]; \, \exists \,  {\bf v} \in \tilde {\cal V} \hbox{ \sl s.t.  } {\bf v} = {\bf v}_r  \right \}Ê\end{equation} 

We prove the following. 
\begin{lemme} The space ${\cal V}$ is dense in $V$. 
\end{lemme} 

{\bf Proof.} Let ${\bf v} \in V$. Since $\tilde \Om$ is simply connected and has a Lipschitz boundary, one knows thanks to Corollary 2.5 in \cite{GR86} that 
$\tilde {\cal V}$ is dense in $\tilde V$. Therefore, there exists a sequence  
$({\bf w}_n)_{n \in \N}$ of fields in $\tilde {\cal V}$ that converges to 
  ${\bf v}^e$ in the space $\tilde V$.  One obviously has 
  $$ \int_{\Om} | \g  (({\bf w}_n)_r - {\bf v}) |^2 \le 
  \int_{\Om^s} |\g  ({\bf w}_n - {\bf v}^e) |^2. $$ 
This shows that the sequence $(({\bf w}_n)_r )_{n \in \N}$ converges to ${\bf v}$ in 
  $V$ and each $({\bf w}_n)_r$ lies in ${\cal V}$ by definition. The lemma is proven. 

\subsubsection{The variational Problem}
For the sake of the simplicity, up to now and throughout the paper we shall note 
$\nu_t(k)$ instead of $\nu_t (k, \x)$. Notice firstly that $\forall \, (\vittest_1, \vittest_2) \in {\cal V}^2$, $\forall \, (k,q) \in {\cal D} (\Om)^2$ one has 
$$ -  \int_{\Om} (\g \cdot \boldsymbol {\sigma } (\vittest_1, q, k)) . \vittest_2 = 
- \int_{\G_o} (\boldsymbol {\sigma } (\vittest_1, q, k).  {\bf n} ) .  \vittest_2 + \int_{\Om} 2 \,\nu_t (k) \boldsymbol {\E } \, (\vittest_1) : \boldsymbol {\E } (\vittest_2).$$ 
The variational formulation of the problem is the following, where the pressure does not appear anymore and will be recovered using The De Rham Theorem. In the following, one denotes 
\BEQ \label{100806} W = L^2 ([0,T], V) \cap L^\infty ([0,T], (L^2(\Om))^2 ). \EEQ

{\sc Find 
\begin{eqnarray}Ê&& \bof  \in W, \quad \bof( 0, \x) = \vit _0 (\x) - \vittest_0 (\x) \quad \hbox{\sl a.e in } \, \Om\\
&& k \in L^\infty ([0,T], L^1 (\Om)) \cap  ( \bigcap_{p<4/3} L^{p} ([0,T], W^{1,p} (\Om) )
) 
 \end{eqnarray}Ê
with
\BEQ \p_t \bof \in L^{8/5} ([0,T], V') \cap W',  \EEQ
and such that $\forall \, {\bf v}Ê\in  L^2 ([0,T], V)$,  
\begin{equation}\label{VARU}Ê\begin{array} {l} \displaystyle <\p_t \bof, {\bf v} > + \int_0^T \int_\Om (\bof \g ) \bof \, . 
{\bf v} + \int_0^T \int_\Om 2 \, \nu_t (k) \E (\bof) :  \E ({\bf v})  +
\int_0^T \int_{\G_o} {1 \over 2} 
[(\bof + \vittest_0). {\bf n}]^- \bof \, . {\bf v}   \, - Ê \\ \displaystyle \int_0^T \int_{\G_o}
[(\bof + \vittest_0). {\bf n}] \vittest_0. {\bf v} + 
\INT {\cal P} (\bof) . {\bf v} + \INT [(\vittest_0 \g ) (\bof +\vittest_0) \, . {\bf v}  + \\
\displaystyle \INT 2 \, \nu_t (k) \E ( \vittest_0 ) : \E ({\bf v} )  + \int_0^T \int_{\G_o} 
(\boldsymbol {\sigma } (\vittest_0, q_0, k). {\bf n} ) .  \vittest = 0, 
\end{array} 
\end{equation} 
for all  $\displaystyle r \in  {\cal D}' ([0, T]\times \Omega)$, with  $r(T, \cdot) = 0$, 
\begin{equation}Ê\label{VARK} \begin{array} {l} \displaystyle - \int_0^T \int_\Om \p_t r k + 
\int_0^T \int_\Om ((\bof+ \vittest_0) \g ) k \, . 
r + \INT \mu_t (k) \g k :  \g r  + \int_\Om k_0 (\x) r (0, \x) d\x 
=  \\ \displaystyle
\INT [2 \nu_t (k) | \E (\bof) |^2  - {\cal E} (k ) + \nu_t (k) (4  \E (\bof) \E (\vittest_0) 
+ 2 | \E (\vittest_0) |^2 )  ] r 
\end{array} 
\end{equation} 
}

\subsubsection{Consistency of the variational formulation} 
The variational formulation for the $k$-equation is the classical one, as in \cite{RL97}, \cite{RL97B}, \cite{RL06} and \cite{LL06}. The variational formulation for the velocity is also classical up to  the boundary terms. Each boundary term where $\vittest_0$ is involved is nice since $\vittest_0$ does not depend uppon the time and is equal to 
$\vit_{\hbox{\tiny I}}$ on $\G_o$ which is in particular in $L^\infty (\G_i)$. 
Nevertheless the term 
$$\int_0^T \int_{\G_o} [\bof . {\bf n}]^- \bof \, . {\bf v}  $$
is fearsome. We prove the following lemma which guarantees the consistency of the variational formulation above. For the sake of simplicity and as far as no confusion occurs, we still denote by ${\bf v}$ the trace of ${\bf v}$ for any ${\bf v} \in W$. Moreover,  one defines the $W$ norm by 
$$ ||Ê{\bf v} ||_W = || {\bf v} ||_{L^2 ([0,T], V)} + || {\bf v} ||_{L^\infty ([0,T], (L^2 (\Om)^2)}.$$

\begin{lemme} \label{LL12} Let $(\bof, {\bf v}) \in W \times W$. Then 
\BEQ \label{101008} \int_0^T \int_{\G_o} [\bof . {\bf n}]^- \bof \, . {\bf v} \le C || \bof ||_W ^2 || {\bf v} ||_W,
\EEQ 
where $C$ is a constant that only depends on $\alpha$ and $\beta$. Moreover, there also exists a constant $\tilde C$ such that 
\BEQ \label{101108} \forall \, \tilde {\bf v} \in L^{8/3} ([0, T], V), \quad 
\int_0^T \int_{\G_o} 
[\bof . {\bf n}]^- \bof \, . {\bf v} \le C || \bof ||_W ^2 || \tilde {\bf v} ||_{L^{8/3} ([0, T], V)}
\EEQ
\end{lemme} 

{\bf Proof.} Let ${\bf v} \in W$. On starts from the classical interpolation inequality (see in \cite{LM68})
$$ || {\bf v}||_{H^{3/4}} \le ||{\bf v}||_{L^2}^{1/4} || {\bf v} ||_{H^1}^{3/4}  .$$
One deduces that 
\BEQ \label{JNSP} || {\bf v}||_{L^{8/3} (H^{3/4})} \le 
||{\bf v}||_{L^\infty (L^2)}^{1/4} || {\bf v} ||_{L^2(V) }^{3/4} \le 
{1\over 4} ||{\bf v}||_{L^\infty (L^2)} + {3 \over 4}   || {\bf v} ||_{L^2(V) } \le || {\bf v} ||_W. \EEQ
One deduces by the trace Theorem that 
\BEQ \label{JNSP1} || {\bf v}||_{L^{8/3} (H^{1/4} (\G_o))} 
 \le C || {\bf v} ||_W. \EEQ
Moreover, thanks to the Sobolev Theorem, 
\BEQ \label{JNSP2} || {\bf v}||_{L^{8/3} (L^4 (\G_o))} 
 \le C || {\bf v} ||_W. \EEQ
Let $\bof \in W$. It is clear that at $\G_0$, 
$$ \bof . {\bf n} \in L^\infty (H^{-1/2} (\G_0)) \cap L^2 (H^{1/2} (\G_0)).$$
By using again a simple interpolation inequality one deduces easily that 
\BEQ \label{PNSP}
|| \bof . {\bf n} ||_{L^4 (L^2 (\G_o))} \le C || \bof ||_W. \EEQ
Therefore, 
$  (\bof . {\bf n} ) \bof \in L^{8/5} (L^{4/3} (\G_0))$, as well as 
$  (\bof . {\bf n} )^- \bof$ and one has 
\BEQ   \label{NNSP}  || (\bof . {\bf n} )^- \bof ||_{L^{8/5} (L^{4/3} (\G_0))} 
\le C || {\bof } ||_W^2. \EEQ
The rest of the proof is now a direct consequence of 
$(\ref{JNSP} )$, $(\ref{JNSP2} )$ and H\"older inequality.

 \subsection{A priori estimate}

\begin{Proposition} There exists a constant $C_1 = C_1 (\vit_0, u_{\hbox{\tiny I}}, \nu, \alpha, \beta)$ and 
for each $p<4/3$ a constant $C_2 = C_2 (\vit_0, u_{\hbox{\tiny I}}, \nu, \mu, p, \alpha, \beta)$ 
such that for   any smooth solution $(\bof, k)$ to the variational problem $[(\ref{VARK}), (\ref{VARU})]$ one has 
\begin {eqnarray} && \label{EST1} || \vit ||_{L^2([0,T], V)} + 
 || \vit ||_{L^\infty ([0,T], L^2(\Om))} \le C_1, \\
&& \label {EST2} ||Êk ||_{ L^p([0,T], W^{1,p} (\Om))} \le C_2. 
\end {eqnarray}
\end{Proposition}

  {\bf Proof}. We proceed in two steps. We  first estimate  the velocity and then the Turbulent Kinetic Energy (TKE).  
  \medskip
  
  {\sl Step 1. Estimating the velocity. } One multiplies the equation $(\ref{NSH1})$ by ${\bf \tilde{u}}$ and integrates on $\Om$. 
A technical but easy computation using the boundary condition  ${\bf \tilde {u}}$ $(\ref{NSH6})$ yields:

\begin{equation}
\begin{array}{l} \label{INE19}
\displaystyle \frac{1}{2} {d \over dt} 
 ||{\bf \tilde{u}}||^2_{L^2(\Om)} + \int_{\Om} 2 \, \nu_t(k)|\boldsymbol {\E}({\bf \tilde{u}})|^2
+  \int_{\Om} 2 \, \nu_t (k) \boldsymbol {\E}({\bf \tilde{u}}) \boldsymbol {\E}(\vittest_0)+ \\ \displaystyle 
\int_\Om {\cal P} (\bof). \bof - \int_\Om \vittest_0 \otimes (\bof +\vittest_0) : \g \bof + 
{1 \over 2} \int_{\G_0} ((\bof +\vittest_0).{\bf n})^+ | \bof |^2
 = 0.
\end{array}
\end{equation}

Since 
$$0 \le  \int_\Om {\cal P} (\bof). \bof  \quad \hbox{and} \quad  0 \le {1 \over 2} \int_{\G_0} ((\bof +\vittest_0).{\bf n})^+ | \bof |^2, $$
using $(\ref{H1})$ and  $(\ref{REG612})$, the energy equality $(\ref{INE19})$ yields 
\begin{equation}
\label{NRJE}
\begin{array}{lll}
\displaystyle \frac{1}{2} \frac{d}{dt} ||{\bf \tilde{u}}||^2_{L^2(\Om)} + \int_{\Om} 2 \, \nu_t (k) |\boldsymbol {\E}({\bf \tilde{u}})|^2  & \le &  
N \int_{\Om} | \boldsymbol {\E} (\bof) | | \boldsymbol {\E} (\vittest_0) |  + 
|| \vittest_0||_\infty \int_\Om | \bof | | \g \bof | \\
&& \displaystyle + || \vittest_0||_\infty^2 \int_{\Om} | \g \bof | 
\end{array}
\end{equation}
By using Young and Korn's inequalities, one has 
\begin{eqnarray} && \label{TT22} \int_{\Om} | \boldsymbol {\E} (\bof) | | \boldsymbol {\E} (\vittest_0) | \le 
 {1 \over 2 \zeta} \int_\Om | \boldsymbol {\E} (\vittest_0) | ^2 + 
 {\zeta \over 2} \int_\Om  | \boldsymbol {\E} (\bof) |^2, \\
 && \label{TT23} \int_\Om  | \bof | | \g \bof |  \le \left ({1 \over 2 \zeta }+C\right ) \int_\Om | \bof |^2 
 + {\zeta \over 2} C \int_\Om   | \boldsymbol {\E} (\bof) |^2.
\end{eqnarray}
where $\zeta	$ will be fixed later on and $C$ is the constant in the Korn inequality.  Finally, by always  using the Young inequality combined with the Cauchy-Schwarz inequality, 
\BEQ 
|| \vittest_0||_\infty^2 
\int_{\Om} | \g \bof | \le {\alpha \beta \over 2 \zeta} || \vittest_0||_\infty^4 + 
 {\zeta \over 2} \int_\Om   | \boldsymbol {\E} (\bof) |^2. \EEQ
 Thereofore, $(\ref{NRJE})$ combined with  $(\ref{H1})$ yields 
 \BEQ \label{EST0606}
 \begin{array}{lll}
 \displaystyle  {d \over 2 dt}  ||{\bf \tilde{u}}||^2_{L^2(\Om)}  
 + \left (\nu_0 - (\zeta/2) (N + C+1) \right )  \int_\Om  | \boldsymbol {\E} (\bof) |^2  
 &\le&  \displaystyle \left ({1 \over 2 \zeta }+C\right )  ||{\bf \tilde{u}}||^2_{L^2(\Om)} \\\\
 &&\displaystyle  +  {N\over 2 \zeta} \int_\Om | \boldsymbol {\E} (\vittest_0) | ^2 \\\\
 && \displaystyle+ {\alpha \beta \over 2 \zeta} || \vittest_0||_\infty^4
\end{array}
 \EEQ
 We choose $\zeta$ be such that $\left (\nu_0 - (\zeta /2 ) (N + C+1) \right ) = \nu_0/2$. One deduces from 
 $(\ref{EST0606})$ and Gronwall's lemma, combined again with Korn's inequality, the existence of $\tilde C = \tilde C(\vit_{\hbox{\tiny I}}, N, \nu_0, \alpha, \beta, T, \vit_0)$, which 
 blows up in a $e^T$ rate  and such that 
 \BEQ  \label{EST0707} || \bof ||_W =  || \bof ||_{L^\infty( [0,T], L^2(\Om))} + || \bof ||_{L^2 ([0,T], V)} \le \tilde C. \EEQ
\medskip

  {\sl Step 2. Estimating the TKE. }   
Notice first that by using the same arguments as in  \cite{RL97} or in \cite{RL97B}, one can make sure that $k \ge 0 $ {\sl a.e.} as far as we assume $k_0 \ge 0$. The boundary terms does not create any troubles because 
$$ \int_0^T \int_{\G_o} (( \bof + \vittest_0). {\bf n})^- k \, (- k^-) =  
\int_0^T \int_{\G_o} (( \bof + \vittest_0). {\bf n})^- ( k^-) ^2 \ge 0.$$
The other terms are like in the general situation studied in \cite{RL97B} chapter 4. From now and throughout the rest of the paper, one works with $k \ge 0$. 
\medskip

Thanks to $(\ref{EST0707})$, we can use  the Classical Boccardo-Gallou\"et estimate (see \cite{BG89}). By a proof already done in \cite{RL97}, \cite{RL97B}, \cite{RL06} and \cite{LL06} and
since we are working in a 2D case and $k=0$ on 
 $\p \Om$, one deduces  that 
 \BEQ \label {ESTK11} \exists \,  \overline C = \overline C (\vit_{\hbox{\tiny I}}, N, \nu_0, \alpha, \beta, T, \vit_0); \quad 
 || k ||_{L^\infty ([0,T], L^1 (\Om))} \le \overline C, \EEQ 
 and 
 \BEQ \label {ESTK111} \forall \, p < 4/3, \quad \exists \,  \hat C = \hat C(p, \vit_{\hbox{\tiny I}}, N, \nu_0, \alpha, \beta, T, \vit_0); 
 \quad || k ||_{L^p ([0,T], W^{1,p}_0 (\Om))}  \le   \hat C .  \EEQ

\medskip

\subsection{End of the proof of the main Theorem} 

The proof now is the same as in \cite{RL97}, \cite{RL97B}, \cite{RL06} and \cite{LL06}, up to the additional terms due to the extra boundary conditions for the velocity. We construct a sequence of smooth approximated solution $(\bof_n, k_n)_{n \in \N}$ (for instance by troncating the l.h.s of the $k$-equation and using the Galerkin method). 
The trick is to prove the weak convergence in $L^2 ([0,T], V)$ of the sequence $(\bof_n)_{n \in \N}$ (up to a subsequence) 
to $\bof \in W$ which satisfies the formulation $(\ref{VARU})$, and in particular, that can be taken as a test function in $(\ref{VARU})$. ÊOnce this task is finished, the rest is classical and works as in \cite{RL97}, \cite{RL97B}, \cite{RL06} and \cite{LL06} since we already have obtained all the required {\sl \`a priori} estimates. 
\medskip

Let $\vit \in W$. One has, after a part integration on the convective term, 
\begin{equation}\label{VARUn}Ê\begin{array} {l} \displaystyle <\p_t \bof_n, {\bf v} > - \int_0^T \int_\Om \bof_n \otimes \bof_n \g 
{\bf v} + \int_0^T \int_{\G_o} (\bof_n. {\bf n}) \bof_n . \vit + \int_0^T \int_\Om 2 \nu_t (k_n) \E (\bof_n) :  \E ({\bf v})   \\ \displaystyle
+ \int_0^T \int_{\G_o} {1 \over 2} 
[(\bof_n + \vittest_0). {\bf n}]^- \bof_n \, . {\bf v}   \, - Ê  \int_0^T \int_{\G_o}
[(\bof_n + \vittest_0). {\bf n}] \vittest_0. {\bf v} +  \\
\displaystyle
\INT {\cal P} (\bof_n) . {\bf v} + \INT [(\vittest_0 \g ) (\bof_n +\vittest_0) \, . {\bf v}  + \INT 2 \nu_t (k) \E ( \vittest_0 ) : \E ({\bf v} )  +  \\
\displaystyle
 \int_0^T \int_{\G_o} 
(\boldsymbol {\sigma } (\vittest_0, q_0, k). {\bf n} ) .  \vittest = 0, 
\end{array} 
\end{equation}

By using $(\ref{EST0707})$, one knows that the sequence $(\vit_n)_{n \in \N}$ is bounded in $W$ and one may extract a subsequence (still denoted by the same) that weakly converges in $L^2 ([0,T], V)$ and in $L^\infty ([0,T], L^2)$ to some $\bof \in W$. One needs compactness, and for it we shall use the Aubin-Lions Lemma. Of course, all the terms involved in $(\ref{VARU})$ satisfied by each $\bof_n$ are nice except the terms
\BEQ \label {DTER}  \int_0^T \int_{\G_o} (\bof_n. {\bf n}) \bof_n . \vit \quad 
\hbox{and} \quad \int_0^T \int_{\G_o} [\bof_n . {\bf n}]^- \bof_n \, . {\bf v}  \EEQ
which are the worse terms and which constitutes the only new difficulty in this problem compared with previous works already quoted. Thanks to inequality $(\ref{101108})$ combined with $(\ref{EST0707})$, the applications 
$$ {\bf v}  \longrightarrow \int_0^T \int_{\G_o} (\bof_n . {\bf n}) \bof_n \, . {\bf v}  , 
\quad {\bf v}  \longrightarrow \int_0^T \int_{\G_o} [\bof_n . {\bf n}]^- \bof_n \, . {\bf v} $$
are bounded in the space $L^{8/5} ([0,T], V')$ ($(\ref{EST0707})$ holds for the second one, the proof is the same for the first one).
Since we are working in a 2D case, and thanks to the regularity of $\vittest_0$, all the other terms are bounded in 
$L^{2} ([0,T], V')$. Therefore, the sequence $(\p_t \bof_n)_{n \in \N}$ is bounded in 
$L^{8/5} ([0,T], V')$ as well as in $W'$. Applying the Aubin-Lions Lemma, one concludes that 
the sequence $( \bof_n)_{n \in \N}$ is compact in $L^{8/5} ([0,T], (L^2(\Om))^2)$. Hence we are back to the usual situation concerning compactness in this type of problem. We bypass the details. We still denote by $( \bof_n)_{n \in \N}$ the subsequence which converges to $\bof$ almost everywhere in $\Om$ and strongly in 
$L^4 ([0,T], (L^4(\Om))^2)$ (we are the 2D case). 
\medskip

One has analogous compactness properties for the sequence $(k_n)_{n \in \N}$ which converges weakly in each $L^p([0,T], W^{1,p}_0 (\Om))$  (up to a subsequence and $p<5/4$) to some $k $ in the space $ \cap_{p<5/4} 
L^p([0,T], W^{1,p}_0 (\Om))$, almost everywhere in $\Om$ and stronly in 
$L^q ([0,T] \times \Om)$ for some $q>1$. 
\medskip

Passing to the limit in all the terms in $(\ref{VARUn})$ is a classical game and follows proofs done already in previous papers (we are in the 2D case), except concerning the terms $(\ref {DTER})$.  We show how to pass to the limit in the first  one, the second one being treated by the same reasoning. Notice that one has 
$$H^1 \subset H^{3/4} \subset V',$$
the injections being dense and compact. Hence, the sequence $(\bof_n)_{n \in N}$ is compact in the space $L^{8/3} ([0, T], (H^{3/4}(\Om))^2 )$. By uniqueness of the limit, 
it converges to $u$ in this space. 
Following the chain rule of the proof of Lemma \ref{LL12}, one deduces 
that $ (\bof_n.{\bf n})_{n \in N}$ converges strongly to $\bof .{\bf n}$ in 
$L^4 ([0,T], L^2 (\G_0))$ while $(\bof_n)_{n \in \N}$ converges strongly to 
$\bof$ in $L^{8/3} ([0, T], (L^4 (\G_0))^2$. Therefore, 
$$ \lim_{ n \rightarrow \infty} \int_0^T \int_{\G_o} (\bof_n. {\bf n}) \bof_n . \vit = 
\int_0^T \int_{\G_o} (\bof. {\bf n}) \bof . \vit.$$
The rest of the proof is now classical.

\subsection{\label{NTCTKE} Neuman Boundary Condition Type for the TKE} 

We are now working in the case where $k$ does satisfy on $\G_o$
\BEQ \label{BBCC} \mu_t {\p k \over \p {\bf n}} = - (\vit.{\bf n})^{-} k \EEQ
instead of $k=0$. Because this case yields serious mathematical complications, we shall not give a complete proof of the existence result. We shall limit ourself to locating  the difficulties, giving the main {\sl \`a priori} estimate and to indicating the direction to take. Details will be written in a forthcoming paper. 

\subsubsection{Variational Formulation} 
When $k$ satisfies $(\ref{BBCC})$ at $\G_o$ instead of $k=0$, the variational formulation for the k-equation becomes: 
for all  $\displaystyle r \in  C^\infty ([0, T]\times \Omega)$, with $r \vert_{\G_i \cup \G_l} = 0$ and $r(T, \cdot) = 0$, 
\begin{equation}Ê\label{VARKK} \begin{array} {l} \displaystyle - \int_0^T \int_\Om \p_t r k + 
\int_0^T \int_\Om ((\bof+ \vittest_0) \g ) k \, . 
r + \INT \mu_t (k) \g k :  \g r  +  \\ \displaystyle
\int_0^T \int_{\G_o} (( \bof + \vittest_0). {\bf n})^- k \, r + \int_\Om k_0 (\x) r (0, \x) d\x 
=  \\ \displaystyle
\INT [2\nu_t (k) | \E (\bof) |^2  - {\cal E} (k ) + \nu_t (k) (4  \E (\bof) \E (\vittest_0) 
+ 2| \E (\vittest_0) |^2 )  ] r 
\end{array} 
\end{equation} 
The source of difficulty is the additional term 
$$I_k = \int_0^T \int_{\G_o} (( \bof + \vittest_0). {\bf n})^- k \, r.$$  

\subsubsection{\`A priori estimate} 
One starts first with the {\`a priori estimate}. We show that in the following, there is a situation where the Boccardo-Gallou\"et result \cite{BG89} can be applied. 
\medskip

Let $g $ be any non decreasing non negative piecewise $C^1$ bounded function defined on $\R^+$, $G(k) = \int_0^k g(k') dk'$. Notice that 
  $G$ is non negative and thanks to the monotonicity of $g$, one has 
  \BEQ \label {FFF44} \forall \, k \in \R^+, \quad 0 \le  k g(k) - G(k) \EEQ
  Therefore, by choosing $g(k)$ as test function in $(\ref{VARK})$, with $\vit =  \bof + \vittest_0$, one has 
  \BEQ \label{GGGE}  \begin{array} {l} 
\displaystyle {d \over dt} G(k)
 + \int_{\Om} \mu_t(k)| g'(k) |\g k |^2 +  \int_{\G_o} (\vit. {\bf n} )^+ G(k) 
 + \int_{\G_o} (\vit. {\bf n} )^- (k g(k) -G(k) ) =  \\ \displaystyle
 \int_\Om g(k) [2 \nu_t(k) | \boldsymbol {\E}(\vit ) |^2 - {\cal E} (k) ] \end{array}  \EEQ
Since $g$ is non negative, combining 
$(\ref{REG612})$, $(\ref {EST0707})$ and $(\ref {FFF44})$ one has 
\BEQ \label{GGGE2} 
\displaystyle {d \over dt} G(k)
 + \int_{\Om} \mu_t(k)| g'(k) |\g k |^2  \le \tilde { \tilde C} || g ||_\infty, \quad  
 \tilde { \tilde C} = \tilde {\tilde C} (\vit_{\hbox{\tiny I}}, N, \nu_0, \alpha, \beta, T).\EEQ
Therefore one can deduce that the results in \cite{BG89} apply. Therefore 
the estimates $(\ref{ESTK11})$ and $(\ref{ESTK111})$ still hold in this case. 

\subsubsection{Consistency of the variational formulation}

As said already, the difficulty is due to the term $I_k$. Recall that from the proof of Lemma \ref{LL12}, $(( \bof + \vittest_0). {\bf n})^- \in L^4 ([0,T], L^2 (\G_0))$. On the other hand, by combining the trace therorem with the Sobolev Theorem, it easily checked that 
$ k \in \cap_{p<4/3} L^p ([0,T], L^{p \over p-2} (\G_0))$.
Here the critical case is the space $ L^{4/3} ([0,T], L^{2} (\G_0))$, which is not achieved. Therefore, it is not guarantied that the integral $I_k$ is defined.  
\medskip

The way to go round this difficulty is to renormalize the equation for $k$, as in 
\cite{RL97B} chapter 5 and also in \cite{FM90}. Roughly speaking, one does not take a test function $r$ in the equation, but $r \psi (k)$ for functions $\psi$ having compact support. 
Then $I_k$ becomes 
$$I_{k, \psi} = \int_0^T \int_{\G_o} (( \bof + \vittest_0). {\bf n})^- k \psi (k) \, r, $$
which is defined since $k \psi (k) $ is bounded. Of course, when doing this, new terms appear in the variational formulation. This is now out of the scope of the present paper and will be the subject of a next paper.

\section{Numerical simulations} \label{simu}

Simulations have been performed using the free software Freefem++ (see \cite{HP}). It allows computations of 2D and axisymmetric fluid dynamics by the means of the finite elements method (FEM). 

\begin{figure}[htbp] 
\begin{center}
\includegraphics [scale=1.1] {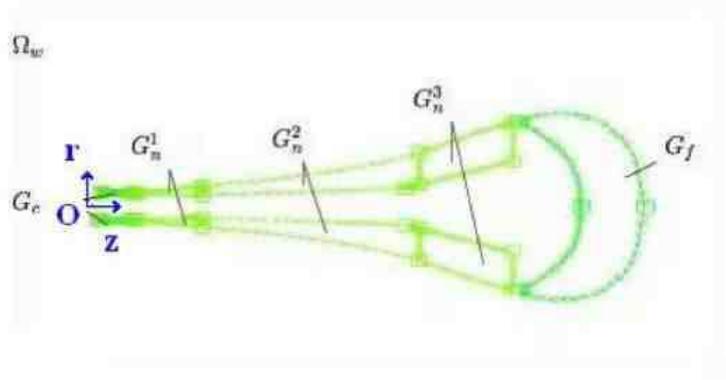} 
\caption{\footnotesize  Geometry of the net}
\label{zoom}
\end{center}
\end{figure}

\medskip
Remember that the net is modeled as a porous membrane and enclosed in a fictive cylinder. Assume that flow is also axisymmetric. Recall this is a strong hypothesis but reasonable in the case of the study of the mean velocity around a rigid net. Then the problem reduces to a 2D one. The geometry shown on Fig. \ref{zoom} and drawn in Freefem++ has an outer net profile and a catch profile in agreement with the model of Boulogne-Sur-Mer.  The inner net profile is defined by the minima of the z component of the velocity located on the LDV profiles (see table \ref{Tabmin}). To take into account the difference of permeability of the net (mainly due to the variations in the mesh opening), the domain $G_n$ has been decomposed in 3 sub-domains: $G_n^1$, $G_n^2$, $G_n^3$. 

\medskip
Let us work in cylindrical coordinates, the z axis being the revolution axis of the membrane:

\begin {equation}
\left \{
\begin{array}{lll}
\displaystyle x & = & r cos \theta, \\
\displaystyle y & = & r sin \theta, \\
\displaystyle z & =& z.
\end{array}
\right.
\end{equation}

Let $\Omega=\{(r,z, \theta), r \in [r_{min}, r_{max}], z \in [z_{min}, z_{max}], \theta \in [0, \pi]\}$.

\medskip
Let $ {\bf u} = (u_r, u_{\theta}, u_z)$ denote the mean velocity unknown in cylindrical coordinates.

\medskip

Assuming a planar flow, then $u_{\theta} = 0$ and thanks to the axisymmetric hypothesis, derivatives with respect to the variable $\theta$ are zero.

\medskip
At a fixed value of $\theta$, we work on a 2D domain: $$\Omega_{r, z}=\{(r,z), r \in [r_{min}, r_{max}], z \in [z_{min}, z_{max}]\} = \Omega_w \cup G_c \cup G_n \cup G_f.$$ Notice we keep the notations: $\Omega_w$ for the fluid domain, $G_c$ for the ring that maintains the model inside the tank, $G_n$ for the membrane (net) domain and $G_f$ for the catch domain. 

\medskip
In the following, the operators (gradient, divergence,...) are considered in cylindrical coordinates.

\medskip
The solid part has a very small permeability, denoted $K_s(r,z) \ll 1$, leading to force the velocity to be zero in that part (then forcing a no slip boundary condition).

\medskip
The porous part has a permeability chosen here to be constant by subdomains $G^n_i$, denoted $K_{G_n^i}$, $i=1, 2, 3$.
\medskip
The fluid domain has an infinite permeability so that the penalization term vanishes in that part, denoted $K_f(r, z) \gg 1$. 

\medskip
The coupled problem $(\ref{NS1}) - (\ref {NS9})$ is implemented under the following variational form.

\subsection{Weak formulation}

At first, let us assume that there is no reflexion at the outer boundary and consider the boundary conditions $(\ref{NS8}) - (\ref {BBCC})$ reduced to:

\begin{eqnarray} 
&&  \label {NS10} \boldsymbol {\sigma}(\vit, p, k). \, {\bf n} \vert_{\G_o} 
=  0, \\
&&  \label {NS11} \displaystyle
{\p k \over \p {\bf n}}  \vert_{\G_o} = 0. 
\end{eqnarray}

\medskip
Moreover, let us replace the no slip boundary condition for the velocity (see equation $\ref{NS7}$) on $\Gamma_l$ by slip boundary condition and the homogeneous Dirichlet condition for $k$ on $\Gamma_l$ by a non homogeneous one:

\begin{equation}
\label{slipBC}
\displaystyle \frac{\partial u_z}{\partial r} = 0,  \quad  u_r = 0, \quad k = k_0 \quad \hbox{ sur } \Gamma_l,
\end{equation}

\medskip
Denote ${\cal V}(\Omega_{r, z})$ and ${\cal Q}(\Omega_{r, z})$ the space defined as:
\begin{equation}
\displaystyle {\cal V}(\Omega_{r, z}) = \{ {\bf v} \in (H^1(\Omega_{r, z}))^2, v_z = 0 \textrm{ sur } \Gamma_i, v_r = 0 \textrm{ sur } \Gamma_i \cup \Gamma_l  \},
\end{equation}
 
\begin{equation}
\displaystyle {\cal Q}(\Omega_{r, z}) = \{ q \in L^2(\Omega_{r, z}) \}.
\end{equation} 

and 
\begin{equation}
\displaystyle {\cal W}(\Omega_{r, z}) = \{ w \in L^2(\Omega_{r, z}), w = 0 \textrm{ sur } \Gamma_i \cup \Gamma_l \}.
\end{equation}

A weak formulation of the coupled problem $(\ref{NS1})-(\ref{NS6})$, with the boundary conditions $(\ref{NS10})-(\ref{NS11})-(\ref{slipBC})$ yields: 

\medskip
\begin{equation}
 \label{FVNSA}
 \left \{
\begin{array}{l}
\textrm{Find $({\bf u} =(u_r, u_z), p, k) \in  {\cal V}(\Omega_{r, z})$ x ${\cal Q}(\Omega_{r, z})$ x ${\cal W}(\Omega_{r, z})$  such that:} \\\\
\displaystyle \int_{\Omega_{r, z}} \frac{\partial {\bf u}}{\partial t} {\bf v} \, |r| \pi drdz + \int_{\Omega_{r, z}} ({\bf u} \nabla) {\bf u} {\bf v} \, |r| \pi drdz - \int_{\Omega_{r, z}} p \nabla \cdot {\bf v} \, |r| \pi drdz  \\\\
\displaystyle + \frac{1}{2} \int_{\Omega_{r, z}} (\nu_0 + \nu_t) (\nabla {\bf u}  + (\nabla {\bf u})^t):(\nabla {\bf v} + (\nabla {\bf v})^t) \, |r| \pi drdz \\\\
\displaystyle + \int_{\Omega_{r, z}} {\cal P} ({\bf u}) (t, (r, z)) {\bf v}\, |r| \pi drdz \\\\
\displaystyle - \int_{\Omega_{r, z}} \nabla \cdot {\bf u} \, q\, |r| \pi drdz = 0, \forall {\bf v} \in  {\cal V}(\Omega_{r, z}), \forall q \in {\cal Q}(\Omega_{r, z}); \\\\
\displaystyle \int_{\Omega_{r, z}} \frac{\partial k}{\partial t} w \, |r| \pi drdz+ \int_{\Omega_{r, z}} {({\bf u} \nabla) k \, w}\, |r| \pi drdz + \int_{\Omega_{r, z}} {\tilde \nu_t} (\nabla k : \nabla w) \, |r| \pi drdz \\\\
\displaystyle - \int_{\Omega_{r, z}} \frac{\nu_t}{2} |\nabla {\bf u} + (\nabla {\bf u})^t|^2 w  \, |r| \pi drdz + \,  \int_{\Omega_{r, z}} \frac{C_3}{\ell(\x)} \, k^{\frac{3}{2}} w \, |r| \pi drdz =0, \\\\
\forall w \in  {\cal W}(\Omega_{r, z}).
\end{array}
\right.
\end{equation}

with

\begin{equation}
\begin{array}{lll}
\displaystyle {\cal P} ({\bf u}) (t, (r,z))& =& \displaystyle ( \frac{1}{K_s(r, z)}( \Ind_{G_f \cup G_c} (r, z)) + \sum_{i=1}^3 \frac{1}{K_{G_n^i}(r, z)} \Ind_{G_n^i} (r, z) \\\\
\displaystyle && \displaystyle+ \frac{1}{K_f(r, z)} \Ind_{\Omega_w}(r, z) ) {\bf u} (t, (r, z))
\end{array}
\end{equation}

\subsection{Finite elements discretization}

\begin{figure}[htbp] 
\begin{center}
\includegraphics [scale=0.8] {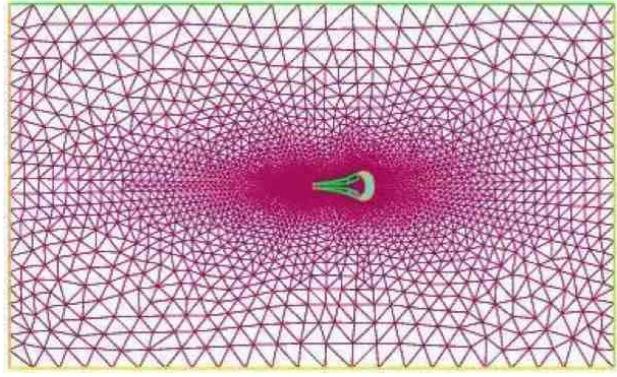} 
\caption{\footnotesize Unstructured mesh of the domain $\Om$ ($10978$ vertices - $21862$ triangles)} 
\label{mesh}
\end{center}
\end{figure}

Using the mesh generator of Freefem++, one builds an unstructured mesh ${\cal T}_h$ of the domain $\{(r,z), r \in [r_{min}, r_{max}], z \in [z_{min}, z_{max}]\}$: $${\cal T}_h= \cup_{i=1,N} K_i.$$ 
Here, $K_i$ are triangle elements. An example of such a mesh, built from the profiles of the different regions is shown on Fig. \ref{mesh}. Recall that the entire domain is meshed even inside the catch and collar regions because equations are set in the entire domain by the means of the permeability of the different media.

\medskip
Mesh refinements are located near the region $G$, since it is the region where most of the turbulence occurs.

\medskip
The space discretization of the problem is based on the finite elements method. The velocity and pressure unknowns are approximated using P2/P1 finite elements. 

The associated discrete finite element spaces are the following:

\begin{equation}
\begin{array}{lll}
\displaystyle {\cal V}_h(\Omega_{r, z})& = &\{ {\bf v}_h =(v_r, v_z) \in ({\cal C}^0(\Omega_{r, z}))^2, \forall K_i \in {\cal T}_h, {\bf v_{h}}|_{K_i} \in P^2(K_i),  {\bf v}_h |_{\Gamma_i \cap \partial K_i} = 0 \\\\
\displaystyle &&\textrm{ et }  v_{r}|_{\Gamma_l \cap \partial K_i} = 0 \},
\end{array}
\end{equation}

\begin{equation}
\displaystyle {\cal Q}_h (\Omega_{r, z}) = \{ q_h \in {\cal C}^0(\Omega_{r, z}), \forall K_i \in {\cal T}_h, q_h \in P^1(K_i) \}.
\end{equation}

The turbulent kinetic energy $k$ is approximated by P2 finite elements. 

\medskip
The associated discrete finite element space is:

\begin{equation}
\begin{array}{lll}
\displaystyle {\cal W}_h(\Omega_{r, z}) & =& \{ w_h \in {\cal C}^0(\Omega_{r, z}), \forall K_i \in {\cal T}_h, w_h \in P^2(K_i), \\\\
\displaystyle &&w_h|_{\Gamma_l \cap \partial K_i} = 0, w_h|_{\Gamma_i \cap \partial K_i} = 0 \}.
\end{array}
\end{equation}

The discrete weak formulation of the problem $(\ref{NS1})-(\ref{NS6})$, $(\ref{NS10})-(\ref{NS11})-(\ref{slipBC})$ is the following:

\medskip
\begin{equation}
 \label{FVNSAD}
 \left \{
\begin{array}{l}
\textrm{Finding $({\bf u}_h, p_h, k_h) \in  {\cal V}_h(\Omega_{r, z})$ x ${\cal Q}_h(\Omega_{r, z})$ x ${\cal W}_h(\Omega_{r, z})$ tel que :} \\\\
\displaystyle \int_{\Omega_{r, z}} \frac{\partial {\bf u}_h}{\partial t} {\bf v}_h \, |r| \pi drdz + \int_{\Omega_{r, z}} ({\bf u}_h \nabla) {\bf u}_h {\bf v}_h \, |r| \pi drdz - \int_{\Omega_{r, z}} p_h \nabla \cdot {\bf v}_h \, |r| \pi drdz   \\\\
\displaystyle + \frac{1}{2} \int_{\Omega_{r, z}} (\nu_0 + \nu_t) (\nabla {\bf u}_h \, + (\nabla {\bf u}_h)^t):(\nabla {\bf v}_h + (\nabla {\bf v}_h)^t) \, |r| \pi drdz   \\\\
\displaystyle + \int_{\Omega_{r, z}} {\cal P} ({\bf u}_h) (t, {\bf x}) {\bf v} \, |r| \pi drdz \\\\
\displaystyle - \int_{\Omega_{r, z}} \nabla \cdot {\bf u} \, q\, |r| \pi drdz  = 0,\forall {\bf v}_h \in  {\cal V}_h(\Omega_{r, z}), \forall q_h \in {\cal Q}_h(\Omega_{r, z}); \\\\
\displaystyle \int_{\Omega_{r, z}} \frac{\partial k_h}{\partial t} w_h \, |r| \pi drdz+ \int_{\Omega_{r, z}} {({\bf u}_h \nabla) k_h \, w_h}\, |r| \pi drdz  \\\\
\displaystyle + \int_{\Omega_{r, z}} {\tilde \nu_t} (\nabla k_h : \nabla w_h) \, |r| \pi drdz - \int_{\Omega_{r, z}} \frac{\nu_t}{2} |\nabla {\bf u}_h + (\nabla {\bf u}_h)^t|^2 w_h  \, |r| \pi drdz  \\\\
\displaystyle  + \,  \int_{\Omega_{r, z}} \frac{C_3}{\ell(\x)} \, k_h^{\frac{3}{2}} w_h \, |r| \pi drdz =0,\forall w_h \in  {\cal W}_h(\Omega_{r, z}).
\end{array}
\right.
\end{equation}


\subsection{Time discretization}

Denote $\delta t$ the time step. Let ${\bf u}^m_h$, $P_h^m$ and $k^m_h$ be the time approximates of the mean velocity, the modified pressure and the turbulent kinetic energy respectively, at the time $t^m = m \delta t$.

\medskip
The convective terms in the problems are approximated using a characteristic Galerkin method \cite{MP}, \cite{HP}. 

\medskip
Consider a convective term like $({\bf u} \nabla) {\bf b}$. 

\medskip
A Taylor expansion of the derivative
\begin{equation}
\label{DP}
\displaystyle  \frac{D{\bf b}}{Dt} = \frac {\partial{\bf b}}{\partial t} + ({\bf u} \nabla) {\bf b}.
\end{equation}

yields the approximation
\begin{equation}
\label{DTA}
\displaystyle  \frac{D{\bf b}}{Dt} \cong \frac {{\bf b}^{m+1} - ({\bf b}^{m}(x-{\bf u}^{m}(x) \delta t)}{\delta t}.
\end{equation}

Let $X(x,t; s)$ be the solution of the problem:
\begin{equation}
\left \{
\begin{array}{l}
\label{EqX}
\displaystyle  \frac{d X}{ds} =   {\bf u}(X, s) \\\\
\displaystyle X|_{s=t} = x.
\end{array}
\right.
\end{equation}

$X(x,t; s)$ is the position at time $s$ of the particule situated at position $x$ at time $t$.
\medskip

Then:
\begin{equation}
\label{DTA2}
\displaystyle  \frac{D{\bf b}}{Dt} \cong \frac {{\bf b}^{m+1} - {\bf b}^m o X^m}{\delta t}.
\end{equation}

where $X^m(x)=X(x, t^{n+1}; t^n)$.

\medskip
Following \cite{MP}, an implicit scheme (see equation $(\ref{schemaNS})$) is chosen for the Navier-Stokes problem with eddy viscosity and a half-implicit one (see equation $(\ref{schemak})$) for the turbulent closure equation.

 \begin{equation}
\label{schemaNS}
\left \{
 \begin{array} {l}
 \textrm{For all $\displaystyle m = 0, ..., \frac{T}{\delta t}$}, \\\\
\textrm{find $({\bf u}^{m+1}_h, p^{m+1}_h, k_h^{m+1}) \in  {\cal V}_h(\Omega_{r, z})$ x ${\cal Q}_h(\Omega_{r, z})$ x ${\cal W}_h(\Omega_{r, z})$ such as:} \\\\
    \displaystyle \frac{1}{\delta t} \int_{{\cal T}_h} ({\bf u}^{m+1}_h-{\bf u}^m_h o X^m_h) \, {\bf v}_h \, |r| \pi drdz - \int_{{\cal T}_h} p^{m+1}_h \, \nabla \cdot {\bf v}_h \, |r| \pi drdz \\\\
  \displaystyle + \frac {1}{2} \int_{{\cal T}_h} (\nu_0 + C_1 \, l \, \sqrt{k^m_h})(\nabla \, {\bf u}^{m+1}_h + (\nabla \, {\bf u}^{m+1}_h)^t) : (\nabla \, {\bf v}_h + (\nabla \, {\bf v}_h)^t) \, |r| \pi drdz\\\\
\displaystyle +\int_{{\cal T}_h} {\cal P}({\bf {\bf u}}^{m+1}_h) \, {\bf {\bf v}_h} \, |r| \pi drdz - \int_{{\cal T}_h} \nabla \cdot {\bf {\bf u}}_h^{m+1} \, q_h \, |r| \pi drdz \\\\
\displaystyle -  \int_{{\cal T}_h} p_h^{m+1} \, q_h \, 0.0000001 \, |r| \pi drdz = 0, \forall {\bf v}_h \in  {\cal V}_h(\Omega_{r, z}), \forall q_h \in {\cal Q}_h(\Omega_{r, z});

 \end{array}
 \right.
\end{equation}

\begin{equation}
 \label{schemak}
 \left \{
 \begin{array} {l}
 \displaystyle \frac{1}{\delta t} \int_{{\cal T}_h} (k^{m+1}_h-k^m_h o X^m_h) \, w_h \, |r| \pi drdz \\\\
 \displaystyle  \int_{{\cal T}_h} (C_2 \, l \, \sqrt{k^m}) (\nabla k^{m+1} : \nabla w_h) \, |r| \pi drdz \\\\
\displaystyle + \frac{1}{2} \int_{{\cal T}_h} (-C_1 \, l \, \sqrt{k^m})|\nabla \, u^{m}_h + (\nabla \, u^{m}_h)^t|^2 \, \frac{k^{m+1}_h}{k^m_h} w_h \, |r| \pi drdz \\\\
\displaystyle + \int_{{\cal T}_h}  \frac{C_3}{\ell(\x)} \sqrt{k^m_h} k^{m+1}_h w_h \, |r| \pi drdz = 0,
 \end{array}
 \right.
\end{equation}

where $X^m_h (x)$ is a numerical approximation of $X^m(x)$. The parameters $C_i$ (i=1, 2, 3) are adimentionalized constants.

\medskip
The penalization term in equation $(\ref{schemaNS})$  $$\displaystyle  \int_{{\cal T}_h} p_h^{m+1} \, q_h \, 0.0000001 \, |r| \pi drdz$$

leads to a more regular problem \cite{HP}.

\medskip
The initial values for the velocity and pressure unknowns $({\bf u}^{0}_h, p^{0}_h)$  are obtained by solving an auxiliary Stokes problem, and the turbulent kinetic energy  $k_h^{0}$ is initialized to a constant in the entire domain.

\medskip
The solving process is iterative. As soon as the final time $T$ is not reached, one solves numerically the kinetic energy problem, then the Navier-Stokes/Brinkman with eddy viscosity part, the time step is increased, the kinetic energy part is solved again and so on.

\subsection{\label{PARSET} Parameters settings}

Different parameters have to be set to perform the simulations:

\medskip
\quad - the parameter $\ell (\x)$ in the definition of the eddy viscosity function (see equation $(\ref{nut})$) is defined as a constant in each triangle, its value in a triangle being equal to the longest edge of this triangle, 

\medskip
\quad - the water kinematic viscosity  $\nu_0$ in equation (see equation $(\ref{nut})$): $\nu_0 = 1.141*10^{-6}$ $\hbox{m}^2 \hbox{s}^{-1}$ at $15\,^{\circ}\mathrm{C}$,

\medskip
\quad - the initial turbulent kinetic energy equal to a constant in the entire domain and equal to $0.01$ $\hbox{m}^2 \hbox{s}^{-2}$,

\medskip
\quad - the permeability $K$ in the different regions: 

\begin{equation}
\begin{array}{lll}
K(\bf x) & = &
\left \{
\begin{array}{l}
\displaystyle 10^4 \textrm{ $\hbox{s}^{-1}$ in the fluid region $\Om_w$;} \\
\displaystyle 10^{-6} \textrm{ $\hbox{s}^{-1}$ in the catch region $G_c$ and collar region $G_f$;} \\
\displaystyle 1  \textrm{ $\hbox{s}^{-1}$ in the net region $G_n^1$;}\\
\displaystyle 5 \textrm{ $\hbox{s}^{-1}$ in the net region $G_n^2$;}\\
\displaystyle 6  \textrm{ $\hbox{s}^{-1}$ in the net region $G_n^3$.}
\end{array}
\right .
\end{array}
\end{equation}

The unit of $K$ is $[\hbox{s}^{-1}]$ since it is formally the ratio between the kinematic viscosity $[\hbox{m}^2 \hbox {s}^{-1}]$ under a permeability surface $[\hbox{m}^2]$.

\medskip
Simulations have shown that the subdomain $G_n^3$ could be considered as permeable (i.e. as a fluid part). In fact, the mesh opening in $G_n^3$ is so high that the meshes do not disturb the flow.

\medskip
\quad - the time step set equal to 0.667 s,

\medskip
\quad - the adimentionalized constants, found numerically: $C_1$ = 0.1; $C_2$ = 0.05; $C_3$ = 0.03.

\subsection{Numerical results}

Using the parameters defined in the previous section, we use the free software FreeFem++ to compute the fluid problem. Runs were made on a bi-processor Pentium Xeon EM64T 3.2Ghz, with 2Go RAM.

\begin{figure}[!h]
\begin{center}
\includegraphics [scale=0.8]{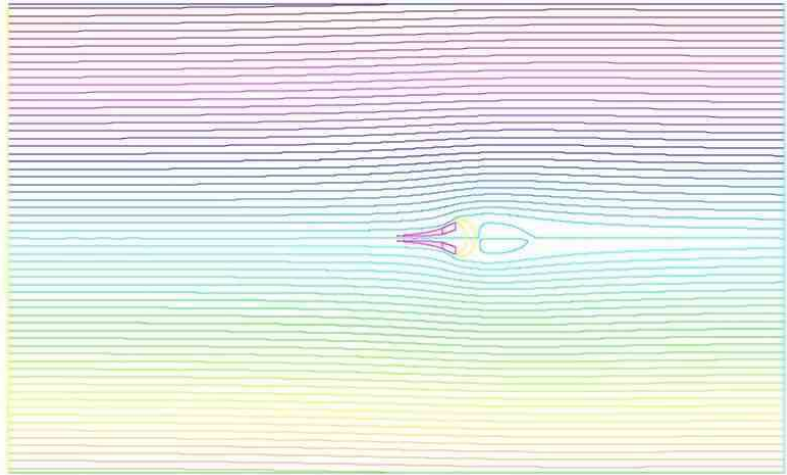} 
\caption{\footnotesize Streamlines} 
\label{streamlines}
\end{center}
\end{figure}

\medskip
The global behavior of the flow is shown on Fig. \ref{streamlines} where the streamlines are drawn.

\begin{figure}[!h] 
\begin{center}
\includegraphics [scale=0.35]{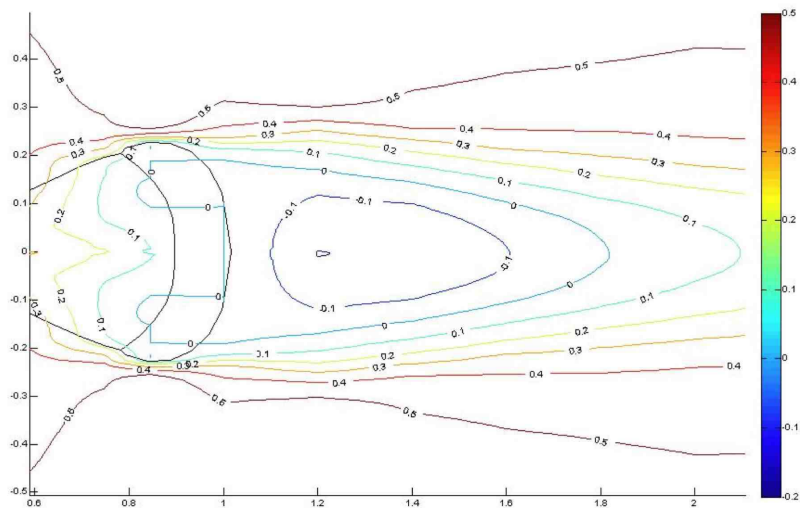} 
\caption{\footnotesize Level curves of $u_z$ behind the catch} 
\label{S18isouz2}
\end{center}
\end{figure}

\medskip
The level curves of the z component of the mean velocity are given in Fig. \ref{S18isouz2}.

\begin{figure}[!h] 
\begin{center}
\includegraphics [scale=0.4]{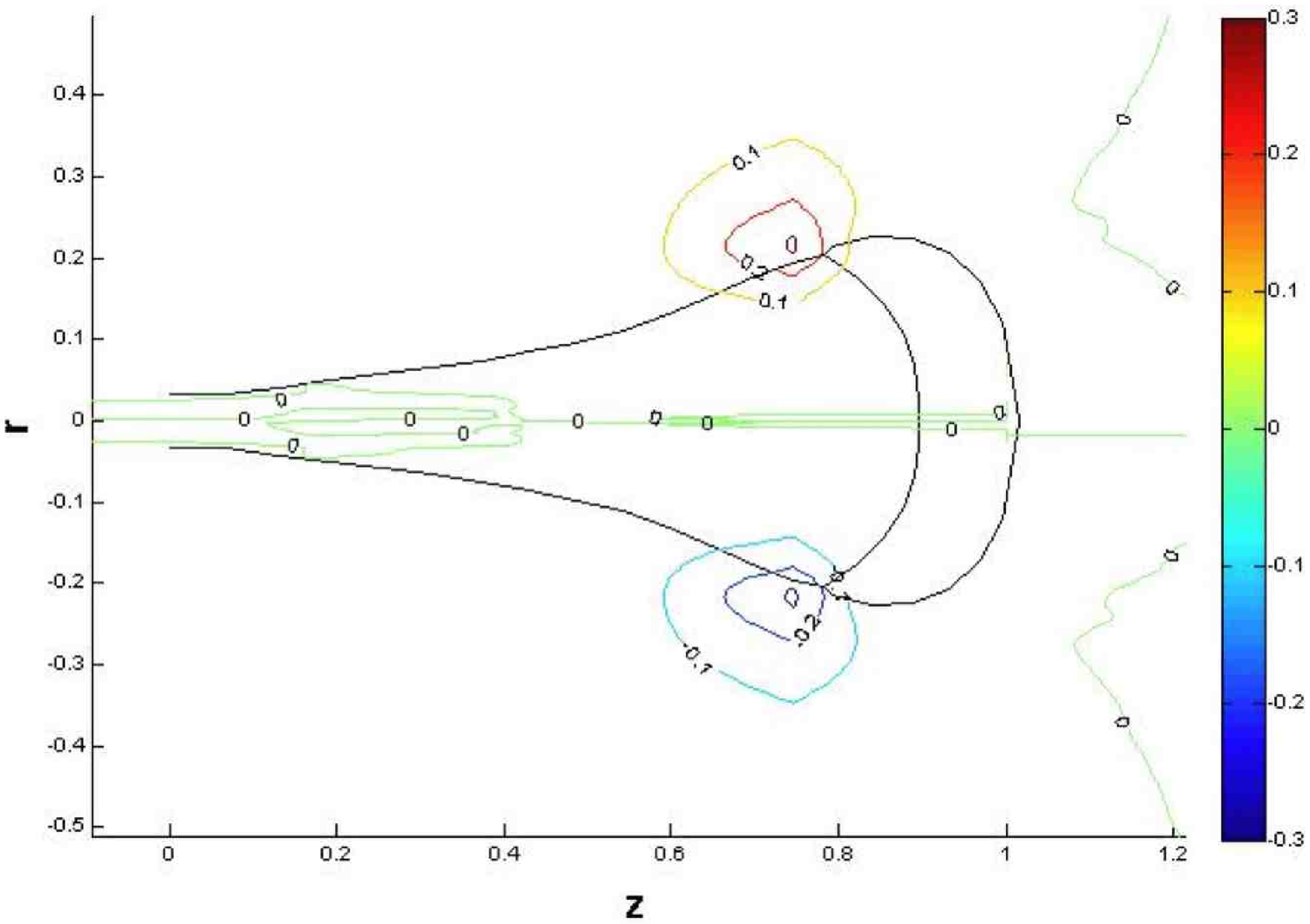} 
\caption{\footnotesize Level curves for $u_r$} 
\label{S18isour}
\end{center}
\end{figure}

\begin{figure}[!h] 
\begin{center}
\includegraphics [scale=0.35]{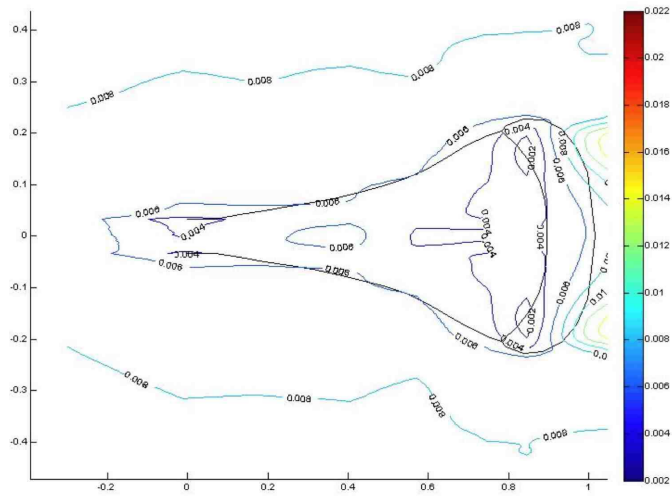} 
\caption{\footnotesize Level curves for $k$ in the surroundings of the net} 
\label{S18isok}
\end{center}
\end{figure}

\begin{figure}[!h] 
\begin{center}
\includegraphics [scale=0.35]{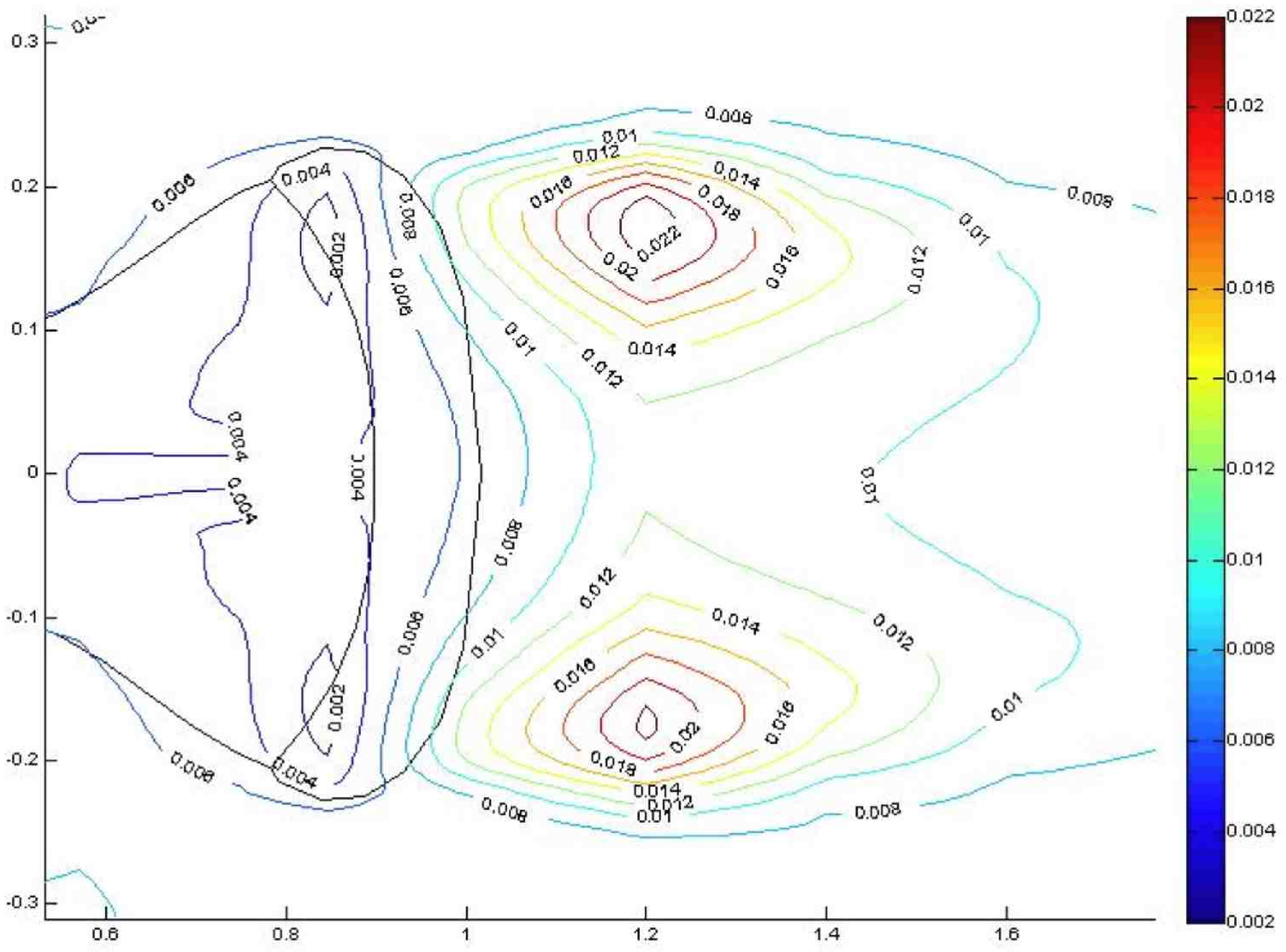} 
\caption{\footnotesize Level curves for $k$ behind the catch} 
\label{S18isok2}
\end{center}
\end{figure}

\medskip
Fig. \ref{S18isour} gives the level curves of $u_r$ and the Fig. \ref{S18isok} and \ref{S18isok2} gives those for the turbulent kinetic energy $k$. 

\medskip
The use of an unstructured mesh leads to a slight asymmetry in the graphics for the turbulent kinetic energy. 

\medskip
Those figures give several results:

\medskip
\quad - a laminar flow at the output (see Fig. \ref{streamlines}). It allows us to keep the simplified boundary conditions $(\ref{NS10})-(\ref{NS11})$ at the output.

\medskip
\quad - the escapement of the inner velocity inside the net takes place just in front of the catch (see Fig. \ref{S18isour}),

\medskip
\quad - the turbulence is mainly located behind the catch and is low in the surroundings of the net (see Fig. \ref{S18isok} and \ref{S18isok2}),

\medskip
\quad - two main eddies are located behind the catch (see Fig. \ref{streamlines}, \ref{S18isouz2}, \ref{S18isok2}).

\begin{figure}[!h]
\includegraphics [scale=0.37]{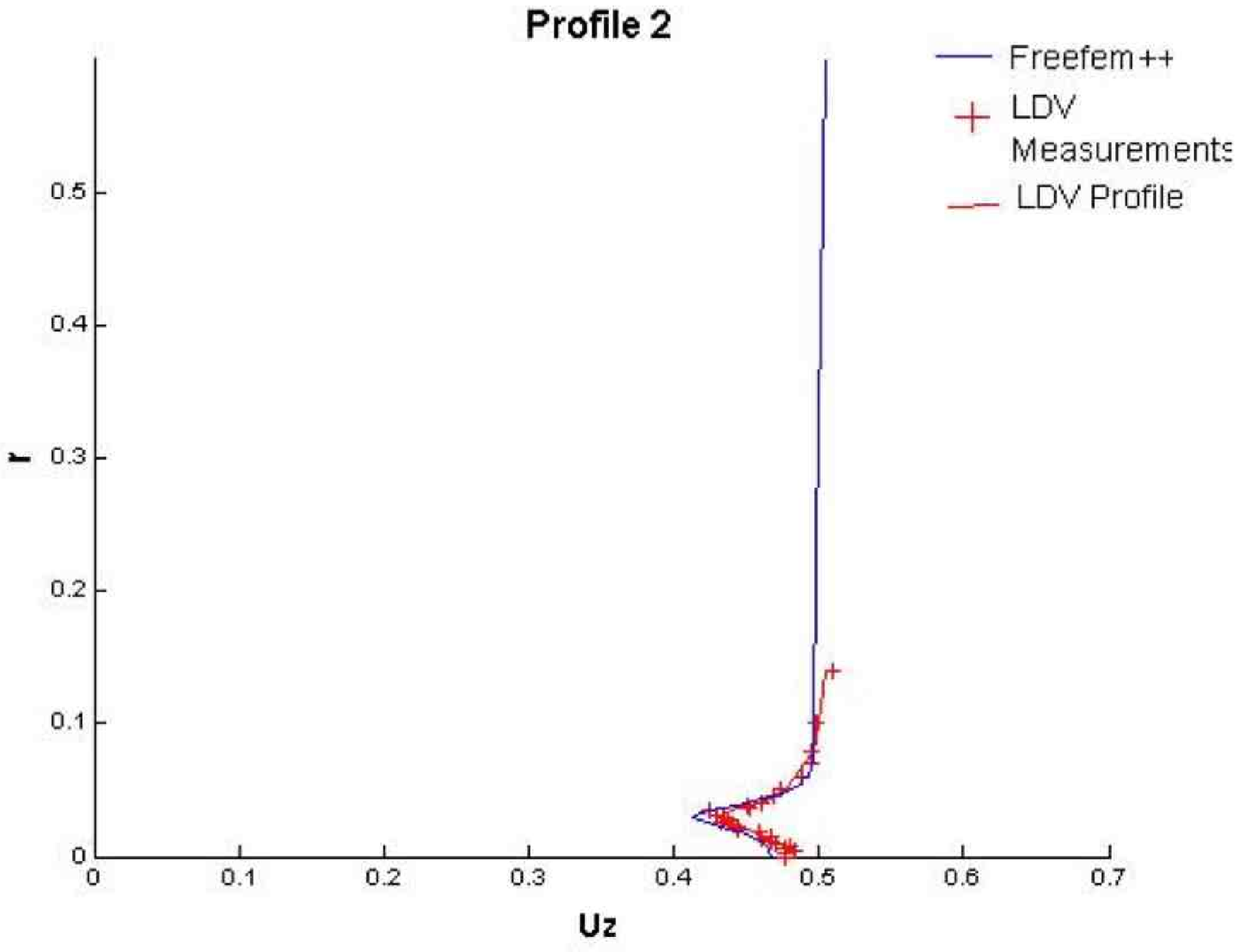} 
\smallskip
\includegraphics [scale=0.37]{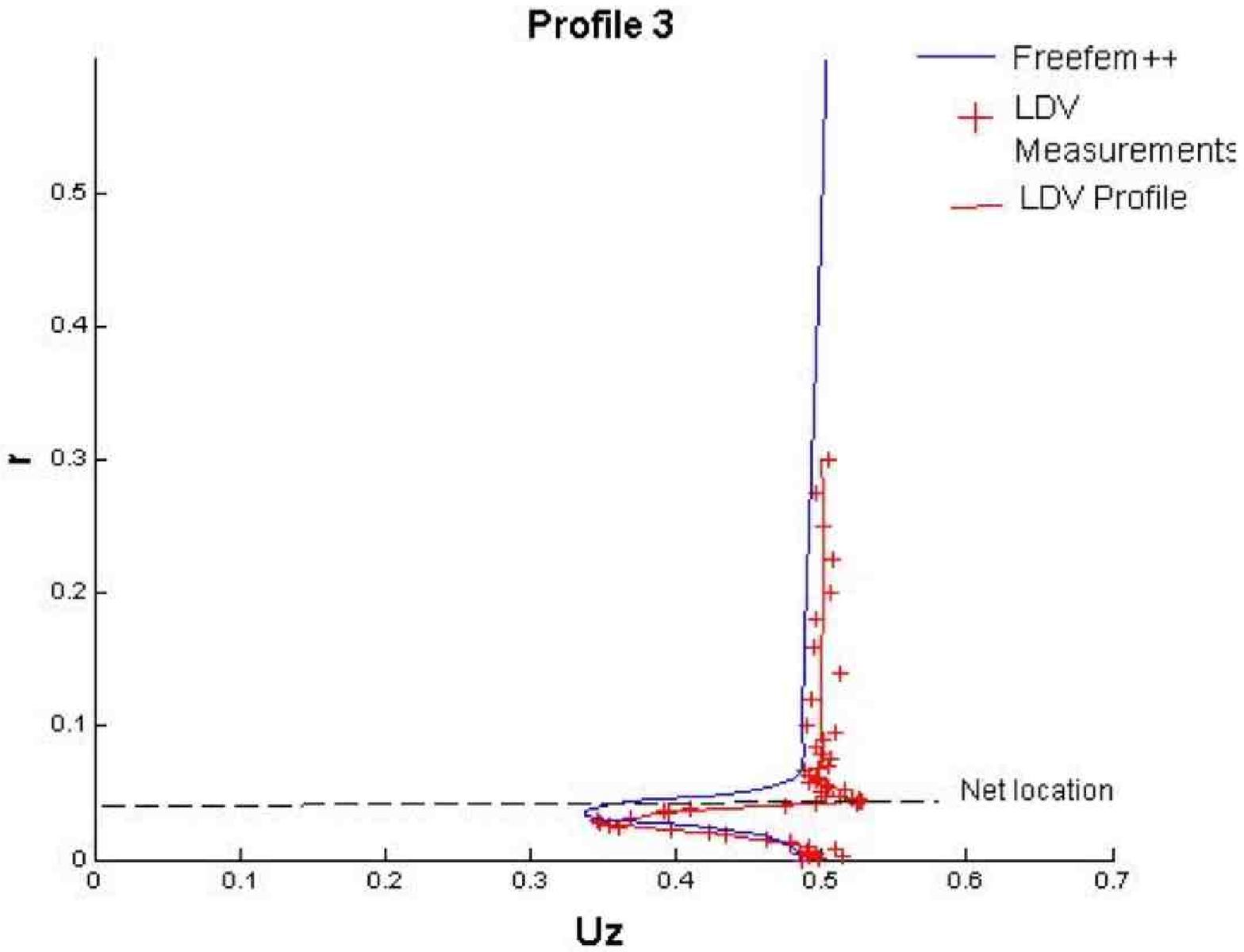} 
\caption{\footnotesize Profiles 2 and  3 after 50 iterations} \label{S18P23} 
\end{figure}

\begin{figure}[!h]
\includegraphics [scale=0.37]{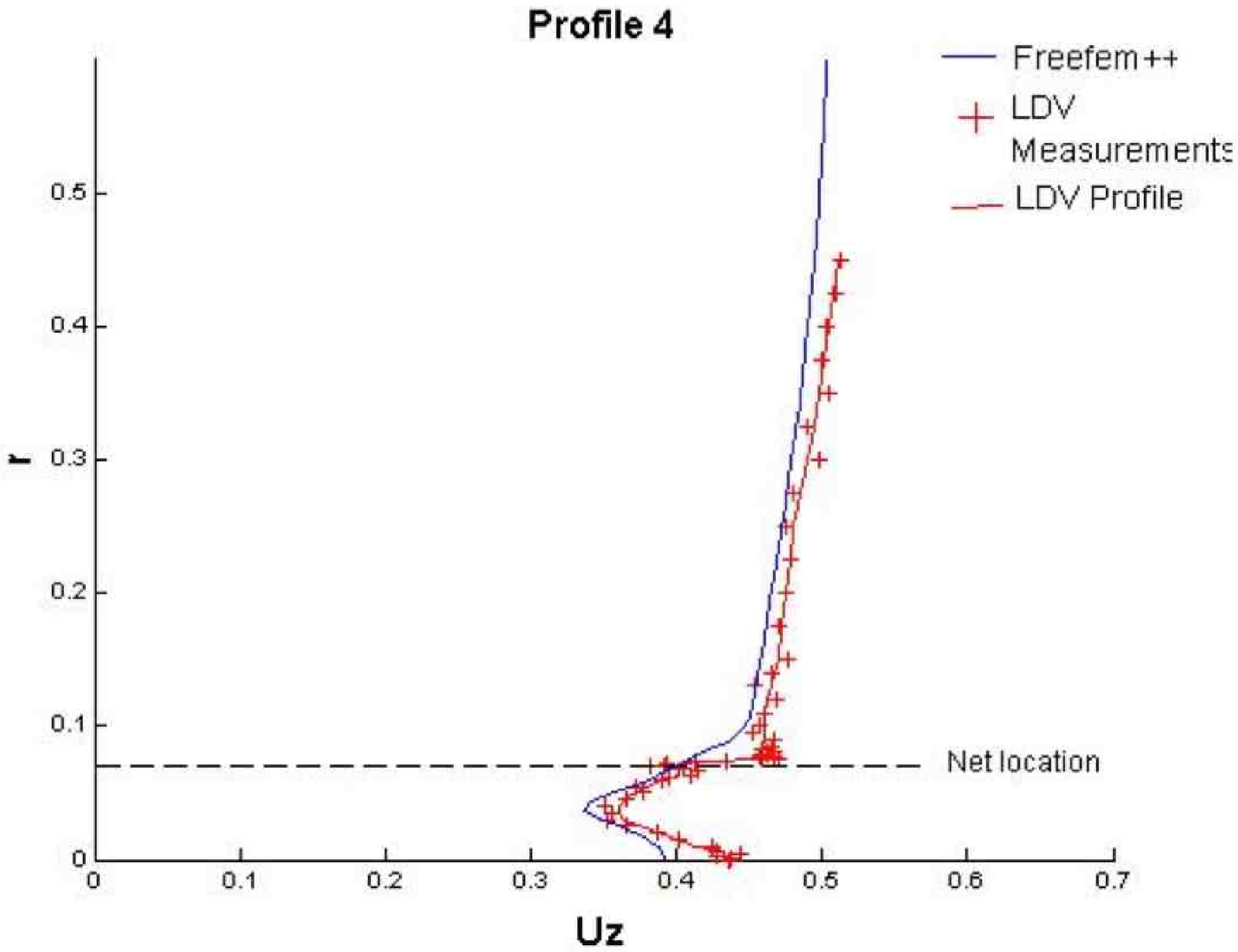} 
\smallskip
\includegraphics [scale=0.37]{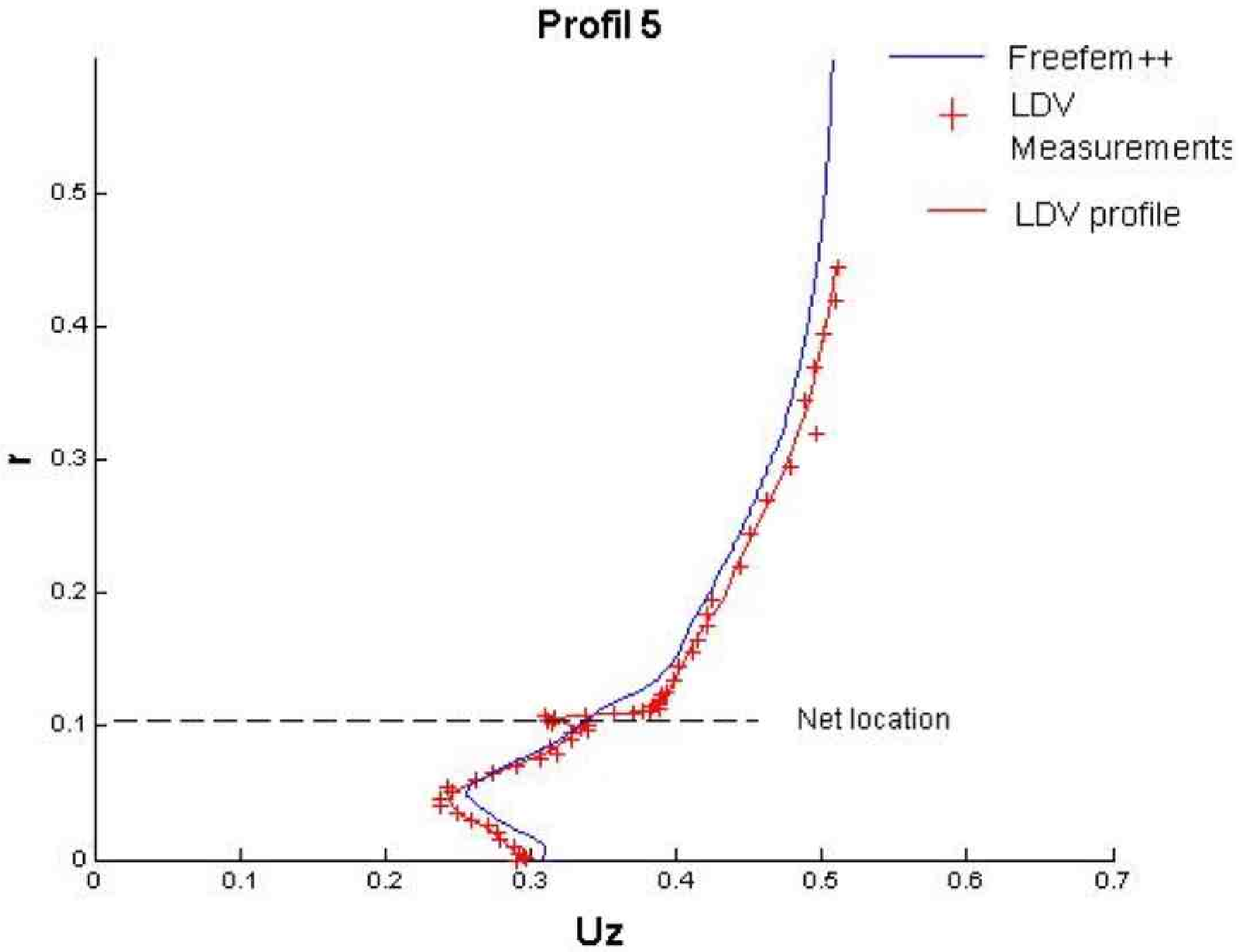} 
\caption{\footnotesize Profiles 4 and  5 after 50 iterations}\label{S18P45} 
\end{figure}

\begin{figure}[!h]
\includegraphics [scale=0.37]{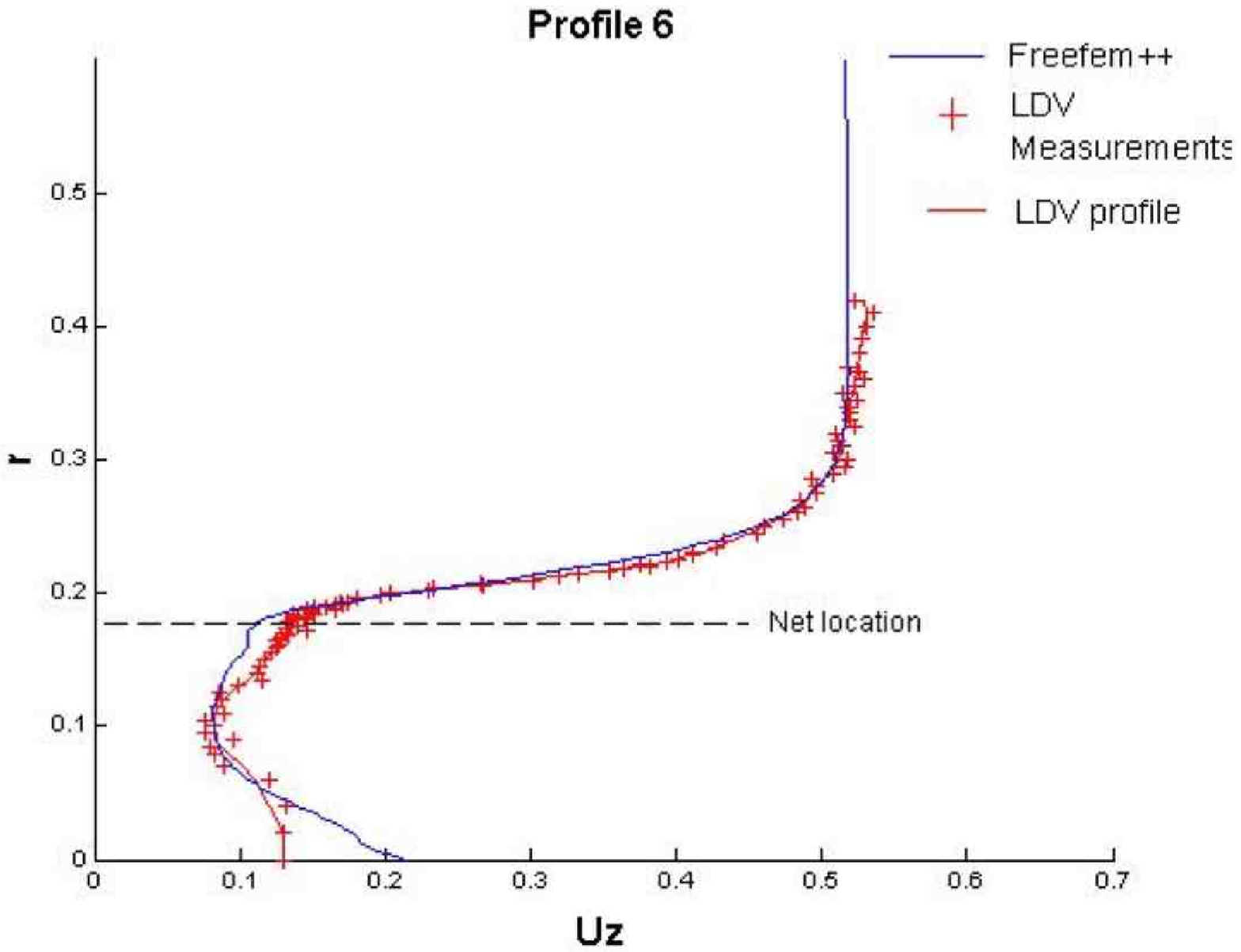} 
\smallskip
\includegraphics [scale=0.37]{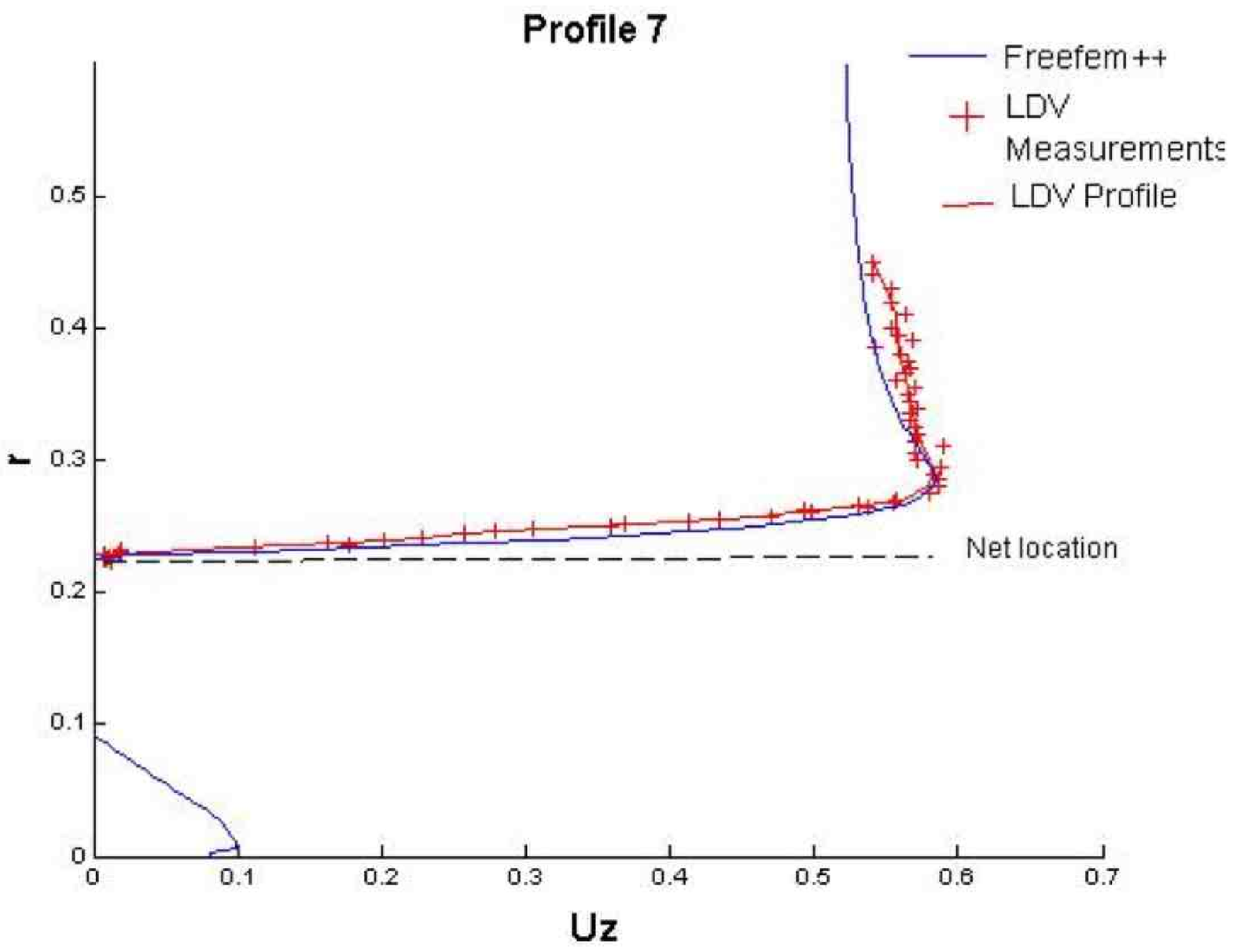} 
\caption{\footnotesize Profiles 6 and  7 after 50 iterations}\label{S18P67} 
\end{figure}

\medskip
Let us compare now the experimental profiles given at the beginning (Fig. \ref{zcomp}), measured by a LDV technique for $u_z$ with those obtained numerically (see Fig. \ref{S18P23}-\ref{S18P45}-\ref{S18P67}). 
\newpage

\begin{figure}[!h]
\includegraphics [scale=0.37]{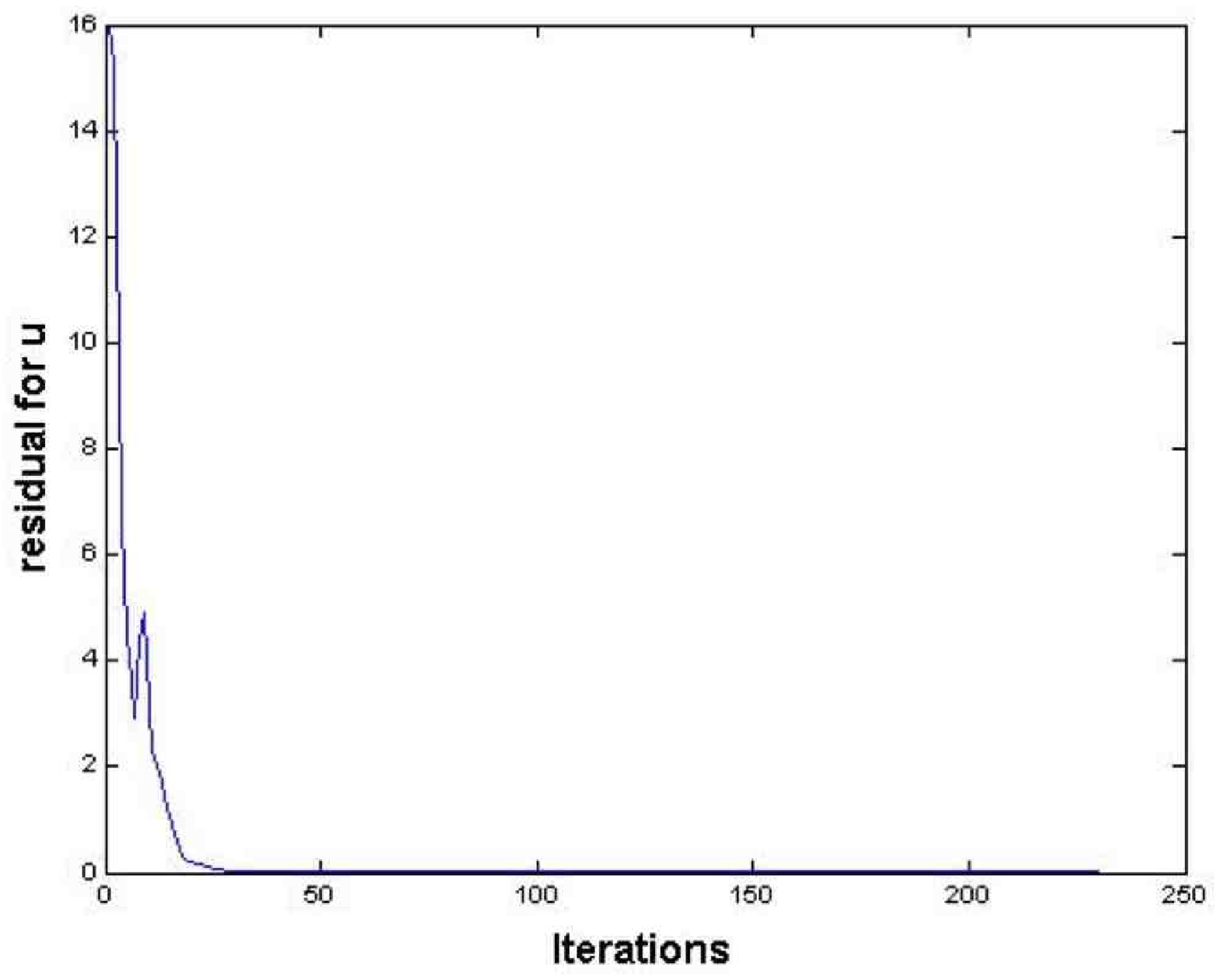} 
\smallskip
\includegraphics [scale=0.37]{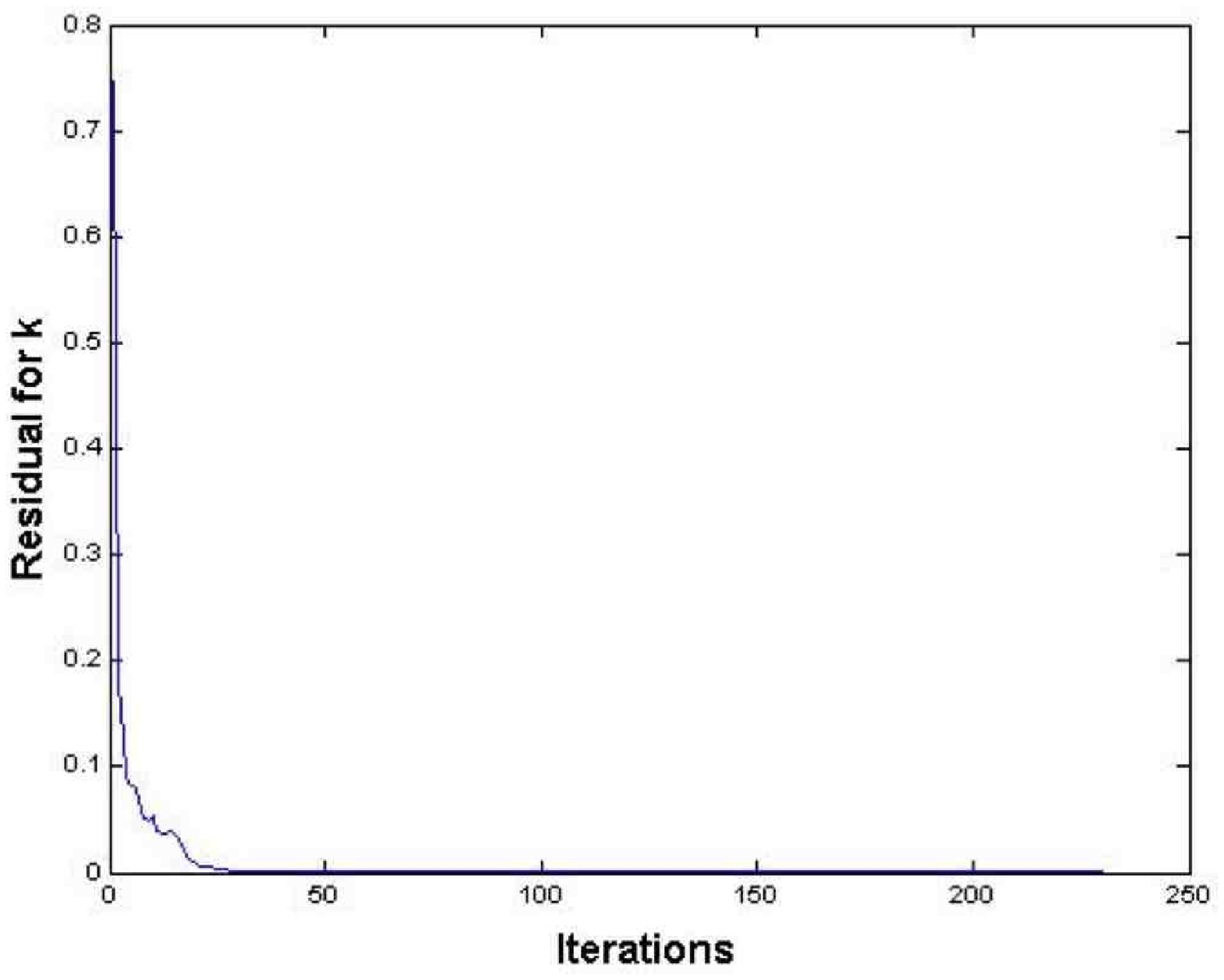} 
\caption{\footnotesize Residual computed for the velocity and TKE vs iterations}\label{res} 
\end{figure}

\medskip
One can see that the numerical profiles fit well with those obtained experimentally (see Fig. \ref{S18P23}-\ref{S18P45}-\ref{S18P67}). 

\medskip
An interesting feature is emphasized by computing the norm 2 of the difference of the velocity and the turbulent kinetic energy between two successive iterations (see Fig. \ref{res}). A stationary state is reached after about 50 iterations: the residual for ${\bf u}$ is equal to 0.00109346, and the one for $k$ equal to 0.000406185. This is in agreement with the fact that we are studying mean quantities.

\medskip
To conclude, we have a model that leads to remarkable results in comparison with the available experimental data. In this particular case of a rigid net, our model looks appropriate. Moreover, this model has the advantage that its application to a 3D problem is possible, especially if we make use of a fictitious domain technique that does not require a complex mesh generation.

{\bf Acknowledgements:} We thank IFREMER and the region Bretagne for the financial support of this work. We are very grateful toward F. Hecht for fruitful discussions about the numerical simulation of our problem. We also thank G. Germain and his team of the IFREMER center of {\it Boulogne-sur-Mer} for the experimental data, B. Vincent of the IFREMER center of {\it Lorient} (France) for his relevant remarks and D. Priour of the IFREMER center of {\it Brest} (France). Finally we express our grateful gratitude to D. Madden who has reread this paper and corrected the english. 

\appendix

\end{document}